\documentclass[12pt]{article}
\usepackage[left=2.5cm,top=2.5cm,right=2.5cm,bottom=1.5cm]{geometry}
\usepackage{graphicx}
\usepackage{latexsym}
\usepackage{epstopdf}
\usepackage{amsmath}
\DeclareGraphicsExtensions{.pdf}
\usepackage{cite}
\begin{document}
	\begin{center}
	\large{\bf{Barrow agegraphic dark energy}} \\
	\vspace{10mm}
	\normalsize{ Umesh Kumar Sharma$^1$, Gunjan Varshney$^2$, Vipin Chandra Dubey$^3$ }\\
	\vspace{5mm}
	\normalsize{$^{1,2,3}$Department of Mathematics, Institute of Applied Sciences \& Humanities, GLA University,
	Mathura-281406, Uttar Pradesh, India \\
	\vspace{2mm}
	$^1$ E-mail:sharma.umesh@gla.ac.in\\
	\vspace{2mm}
	$^2$E-mail:gunjanvarshney03@gmail.com \\
	\vspace{2mm}
	$^3$E-mail:vipin.dubey@gla.ac.in}\\
		
	\end{center}
	\vspace{10mm}
	\begin{abstract}
	
	In this work, we propose a new dark energy model by applying the Barrow entropy and the holographic principle, with a time scale as IR  cut off. Analysing the conformal time as well as universe's age  as infrared cut-offs, we explore the cosmological importance of the suggested dark energy models and examine the universe evolution filled with the proposed DE applicants and a pressure-less matter. We observe that the equation of state, deceleration, the density parameters can present adequate nature, and these models may also explain the late-time acceleration though, the proposed models are unstable except some values of Barrow exponent $\Delta$. Furthermore, we mention the consequences of the presence of  interaction among the universe sectors.

	\end{abstract}

	
	\section{Introduction}

An alternative scenario arises from the holographic principle to describe the dark energy (DE) quantitatively named holographic dark energy (HDE) \cite{ref2}. One can obtain vacuum energy of the holographic principle by commencing the association among the highest length of quantum hypothesis by its ultraviolet cut-off \cite{ref7} that form dark energy at cosmological scales \cite{ ref8, ref9}. The HDE witnesses to direct to fascinating cosmological nature both at its extensive \cite{ref20}, as well as at its modest \cite{ref10} variants, and it also compiles with observational data \cite{ ref47}. The $\Lambda$ model yields some obstacles like the cosmic coincidence and fine-tuning issues and is compatible with observations \cite{ref54a}. Because of these obstacles, the physicists get motivated to study for other DE applicants. Depending on the uncertainty connection of QM (quantum mechanics) \cite{ref21}, ADE (agegraphic dark energy) is an option to the $\Lambda$CDM model.  After discussion the uncertainty in time $t$ is $\delta t = \beta t_{p}^{2/3} t^{1/3}$ for Minkowskian space-time, here $t_{p}$ resembles the reduced Plank time \cite{ref21} and  $\beta$ shows the dimensionless constant of order unity. Wei and Cai introduced a new agegraphic DE model \cite{ref64}, and they described the conformal time $\eta$ as $ a d\eta=dt $ because of facing some difficulties in the original ADE \cite{ref21}, where they took $t$ as the cosmic time rather than the age of the universe. It deserves to mention that for FRW universe,  $\eta$ fulfils the relation $ds^{2} = dt^{2} - a^{2} dx^{2} = a^{2} (d\eta^{2} - dx^{2})$. Each alteration to the entropy of the system modifies the agegraphic DE model considering that the entropy relation has a vital role in this appearance \cite{ref21, ref64, ref54b}. In the literature \cite{ref54c, ref54d}, the ADE models have been studied. \\

At the cosmological frame, the essential step in the holographic principle's statement is that the universe horizon entropy (that is the longest distance) is proportionate to its area and likewise to the black-hole (BH) Bekenstein- Hawking entropy. Though, Covid-19 virus cases motivated the newly Barrow and presented that quantum-gravitational conclusions might propose fractal, complex characteristics for BH formation. This composite system points to restricted volume with the infinite (or finite) area, and consequently formed a distorted BH entropy \cite{ref55}

\begin{eqnarray}
\label{eq1}
S_{B} = \left(\frac{A}{A_{0}}\right)^{1+\frac{\Delta}{2}},
\end{eqnarray}

where $A_{0}$ is the Planck area and $A$  the usual horizon area.  The quantum gravitational perturbation is represented by the
	new exponent $\Delta$. There are some characteristic values for $\Delta$.
	For example, when $\Delta$ = 0 we have the simplest horizon construction. In this case we obtain the well known Bekenstein-Hawking entropy. On the other hand, when $\Delta$ = 1 we have
	the so-called maximal deformation. Therefore, although $\Delta$ is
	not, obviously, a quantum quantity, it is the ``quantum'' effect
	present in Barrow’s entropy expression \cite{ref65}. Marked that with logarithmic corrections, the above quantum gravitational simplified entropy is distinct from the usual "quantum-corrected" entropy \cite{ref56, ref57}, although the involved physicsl principles and foundations  are totally different, despite it relates Tsallis non-extensive entropy \cite{ref58, ref59}. E N. Saridakis proposed the Barrow holographic dark energy (BHDE) model by using Barrow entropy and holographic principle considering the future event horizon as IR cut-off\cite{ref60a}. Very recently, Srivastava and Sharma \cite{ref60aa} constructed the Barrow holographic dark energy using Hubble horizon  as IR cut-off and explored the important cosmological parameters. While the inequality of standard holographic dark energy is presented by $\rho_{D} L^{4} \geq S$, here $L$ shows the length of the horizon and beneath the command $S \propto A \propto L^{2}$ \cite{ref8}, the application of Barrow entropy Eq. (\ref{eq1}) will reach as

\begin{eqnarray}
\label{eq2}
\rho_{D} =  C L^{\Delta-2}.
\end{eqnarray}

With dimensions $[L]^{-2-\Delta}$, $C$ is a parameter. The above mentioned expression gives the standard  holographic dark energy $\rho_{D} = 3 c^{2} M^{2}_{p} L^{-2}$  (where the plank mass is $M_{p}$ ), here  $C = 3 c^{2} M^{2}_{p}$, with the model parameter $c^{2}$. Nevertheless, the BHDE will deviate by standard one and reaches to  various cosmological nature in the situation  when deformation effects quantified by $\Delta$. Applying the Barrow entropy with ``gravity-thermodynamics'' procedure,  Saridakis \cite{ref60aaa} presented the
	 modified cosmological equations which contain new extra terms, and which coincide
	with the usual Friedmann equations in the case where the new Barrow exponent acquires
	its usual value $\Delta$ = 0. In the general case these new terms constitute an effective dark
	energy sector leading to interesting phenomenological behavior, while in the special case
	$\Delta$ = 1 $\Lambda$CDM concordance model is recovered. Here, we are going to apply the Barrow entropy \cite{ref55} to form two Barrow agegraphic dark energy (BADE) models by using  the IR cut-off as the age of the universe and conformal time, and also examine their effects on the universe evolution.\\

This paper is  organized as follows: In Sec. 2, we study BADE model in which the IR cutoff is taken as the universe age and investigate the universe evolution with and without interaction. We introduce a new BADE (NBADE) model and study the cosmic evolution considering the conformal
time as IR cutoff with and without interaction in Sec. 3.
 A summary and concluding
remarks are given in last section.

\section{Barrow agegraphic dark energy (BADE) model}

Taking the universe's age as an infrared cut-off, it is described as

\begin{eqnarray}
\label{eq3}
T= \int_{0}^{a} dt = \int_{0}^{a} \frac{da}{Ha},
\end{eqnarray}

here, $H = \frac{\dot{a}}{a}$ and $a$ are the Hubble parameter and scale factor, respectively. Eq. (\ref{eq2}) is used to form the BADE's energy density

\begin{eqnarray}
\label{eq4}
\rho_{D} = C T^{\Delta-2},
\end{eqnarray}
for $\Delta = 0 $, we update the original ADE model of Cai \cite{ref21}.
Additionally, the present observations reveal a mutual interaction among the DE and DM \cite{ref60b, ref60c, ref60d}. A decision disintegrates the  conservation law for energy-momentum  as
\begin{eqnarray}
\label{eq4a}
\dot{\rho_{m}} + 3 H \rho_{m} = Q,
\end{eqnarray}

\begin{eqnarray}
\label{eq4b}
\dot{\rho_{D}} + 3 H (1+\omega_{D}) \rho_{D} = -Q.
\end{eqnarray}

Here, $\rho_{m}$ and $\rho_{D}$ are DM and DE energy densities respectively. $\omega_{D} = \frac{p_{D}}{\rho_{D}}$
, where $\rho_{D}$ resembles the  dark energy pressure, and also known as the EoS parameter of dark energy. For $Q > 0 (Q < 0)$, there will be an energy transfer from DE (DM) to DM (DE) and such interaction may resolve the coincidence issues \cite{ref62}. Flat FRW universe's first Friedmann equation filled with pressure-less fluid $\rho_{m}$ and BADE ($\rho_{D}$) is formulated as
\begin{eqnarray}
\label{eq5}
H^{2} = \frac{1}{3 m^{2}_{p}} (\rho_{m} + \rho_{D}),
\end{eqnarray}
that can also be revised as
\begin{eqnarray}
\label{eq6}
\Omega_{m} + \Omega_{D} = 1,
\end{eqnarray}
the fractional energy densities can be described as
\begin{eqnarray}
\label{eq7}
\Omega_{m} = \frac{\rho_{m}}{3m^{2}_{p}H^{2}},~~~~~~~~~~
\Omega_{D} = \frac{\rho_{D}}{3m^{2}_{p}H^{2}}.
\end{eqnarray}
Lastly, we simply receive
\begin{eqnarray}
\label{eq8}
r= \frac{\Omega_{m}}{\Omega_{D}} = -1 + \frac{1}{\Omega_{D}},
\end{eqnarray}

for the ratio of the energy densities. One can get an initial calculation for the provided periods of $\Delta$ and $C$ satisfying observation, applying the  values (observational) of the fractional energy densities and $H$, which is visible from Eq. (\ref{eq7}). Applying observational results on other cosmological parameters as the DP (deceleration parameter) and many more, we can get more positive outcomes. Resulting, since we are keen to present the dynamism of this model, estimating $\Omega_{D}^{0} = 0.70$ and $H_{0} = 67$ as the present values of above mentioned parameters \cite{ref61}, we select  values model parameters as $\Delta$ and $C$ forming different nature.

\subsection{ (Q=0) Non-interacting model }

We obtain the EoS parameter by applying the derivative with the time of Eq. (\ref{eq4}) in the conservation equation Eq. (\ref{eq4b}) as

\begin{eqnarray}
\label{eq9}
\omega_{D} = -1 - \frac{\Delta - 2}{3 H T},
\end{eqnarray}
here $T$ presents
\begin{eqnarray}
\label{eq9a }
T = \left(\frac{3 H^{2} \Omega_{D}}{C}\right)^{\frac{1}{\Delta-2}}.
\end{eqnarray}
\begin{figure}[htbp]
	\centering
	\includegraphics[width=8cm,height=8cm,angle=0]{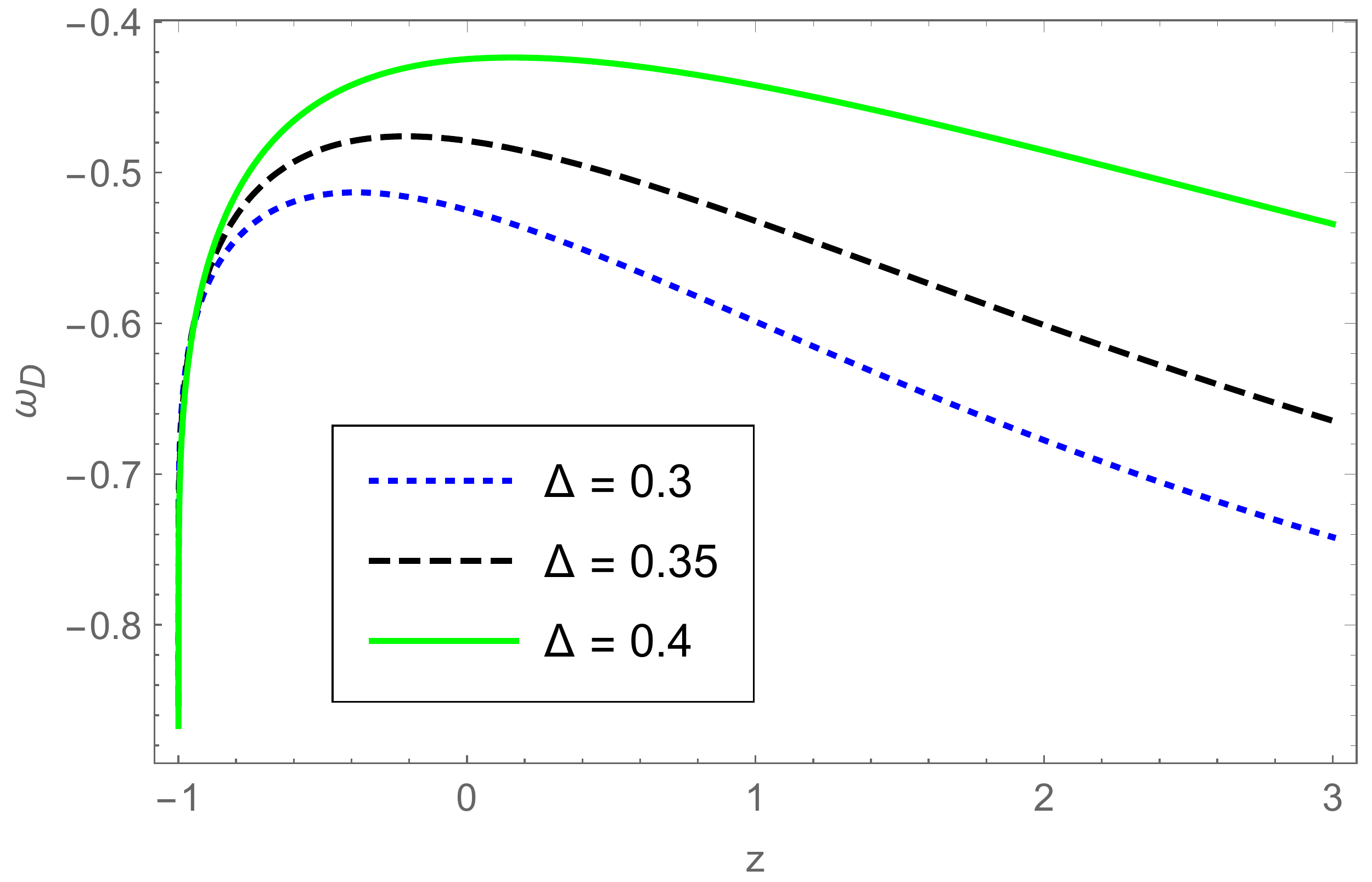}
	\includegraphics[width=8cm,height=8cm,angle=0]{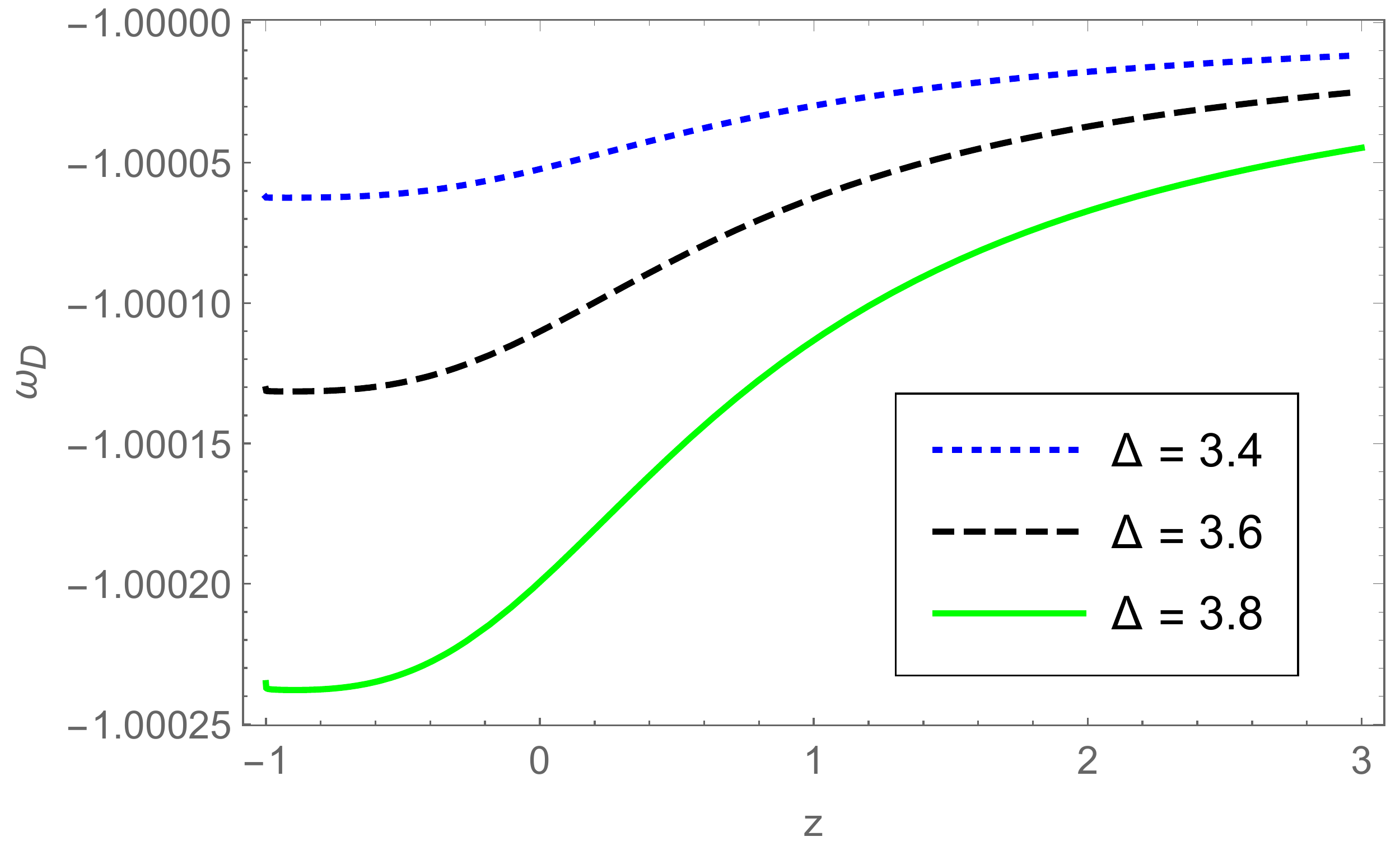}
	\caption{The evolutionary behaviour of equation of state parameter  $\omega_{D}$ versus redshift for non-interacting BADE. Here, we have considered $H_{0}=67$, $\Omega_{D0} =0.70$ and $C=10$.} 
\end{figure}
Besides, by connecting the derivative with time of Eq. (\ref{eq5}) and practising Eqs. (\ref{eq4a}) and (\ref{eq4b}), we grasp
\begin{eqnarray}
\label{eq10}
\frac{\dot{H}}{H^{2}} = - \frac{3}{2} (1-\Omega_{D}) + \frac{\Omega_{D} (\Delta - 2)}{2HT},
\end{eqnarray}
 than can also points to 
 \begin{eqnarray}
 \label{eq11}
 q=-1- \frac{\dot{H}}{H^{2}} =  \frac{1}{2} - \frac{3\Omega_{D}}{2}  - \frac{\Omega_{D} (\Delta - 2)}{2HT}.
 \end{eqnarray}
 \begin{figure}[htbp]
 	\centering
 	\includegraphics[width=8cm,height=8cm,angle=0]{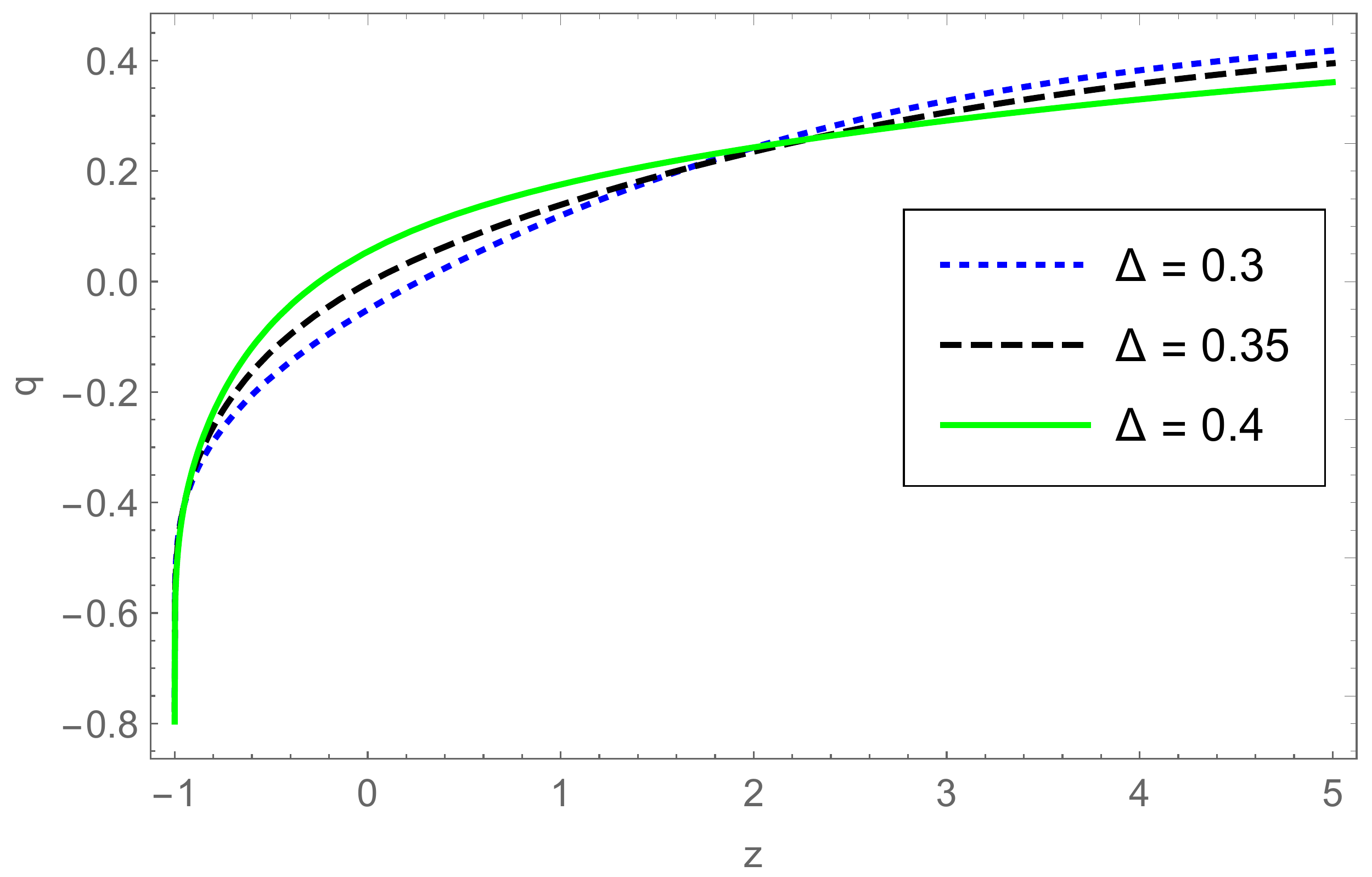}
 	\includegraphics[width=8cm,height=8cm,angle=0]{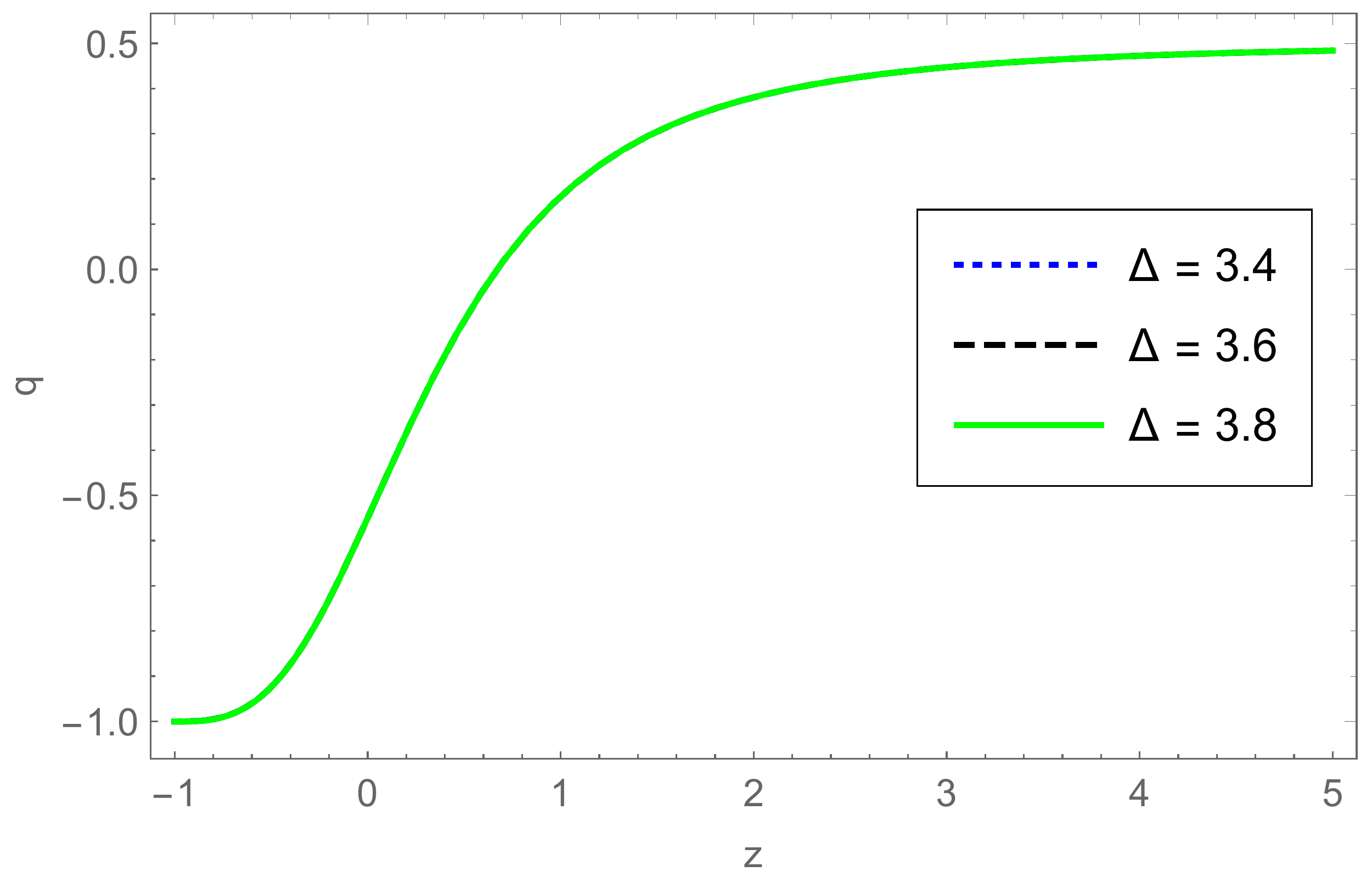}
 	\caption{The evolutionary behaviour of the deceleration parameter $q$ versus redshift for non-interacting BADE. Here, we have considered $H_{0}=67$, $\Omega_{D0} =0.70$ and $C=10$.} 
 \end{figure}
Also a matter of examination to determine $  {\dot\Omega_{D}}$
 \begin{eqnarray}
 \label{eq12}
 {\dot\Omega_{D}} = 2 H \Omega_{D}(1+q) + \frac{\Omega_{D} (\Delta - 2)}{T},
 \end{eqnarray}

here dot signifies the derivative concerning the cosmic time. We calculate $v^{2}_{s}$ (the squared sound speed) to analyse the consequences of disturbances on the standard resistance of the model. 
\begin{figure}[htbp]
	\centering
	\includegraphics[width=8cm,height=8cm,angle=0]{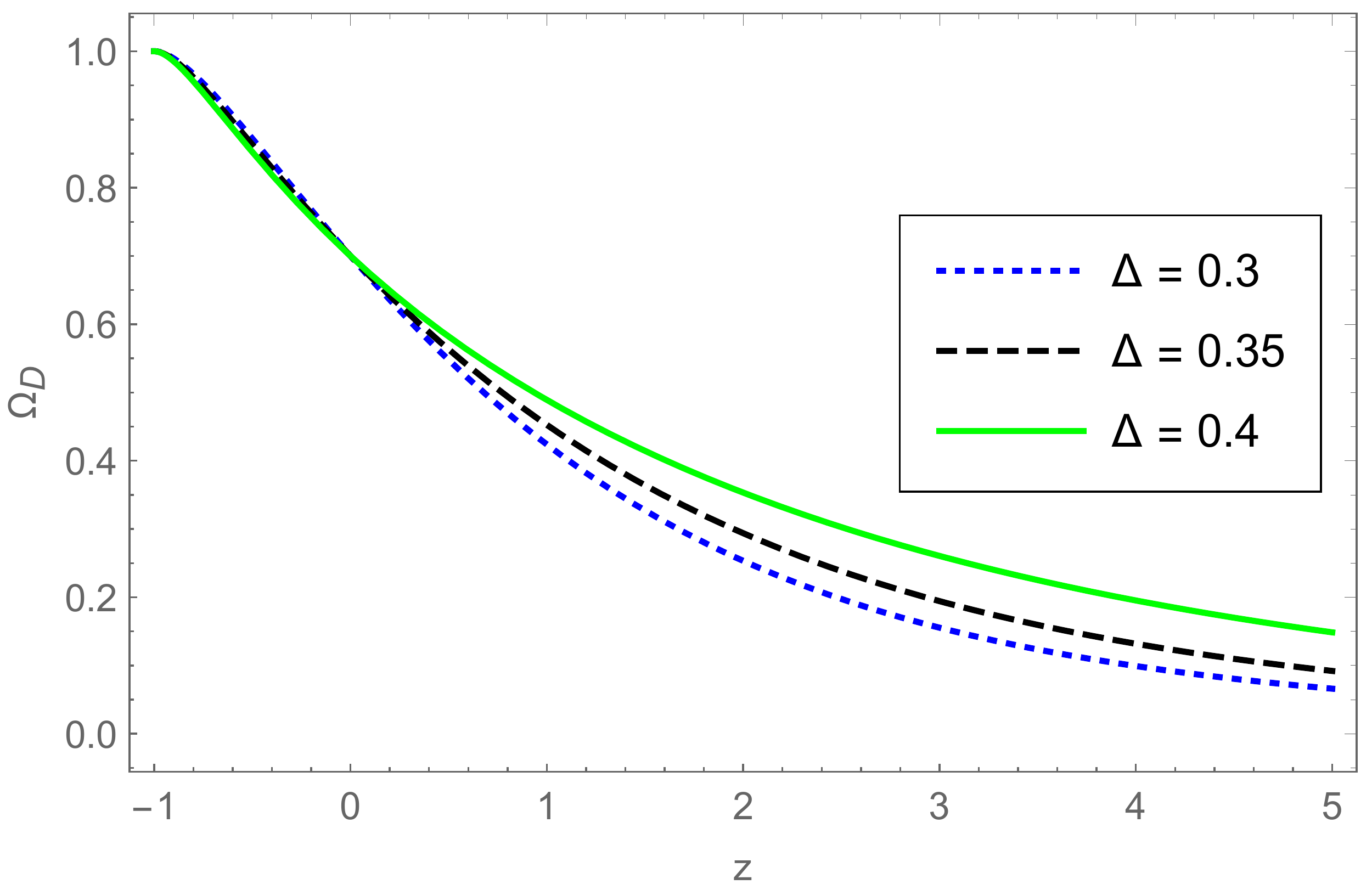}
	\includegraphics[width=8cm,height=8cm,angle=0]{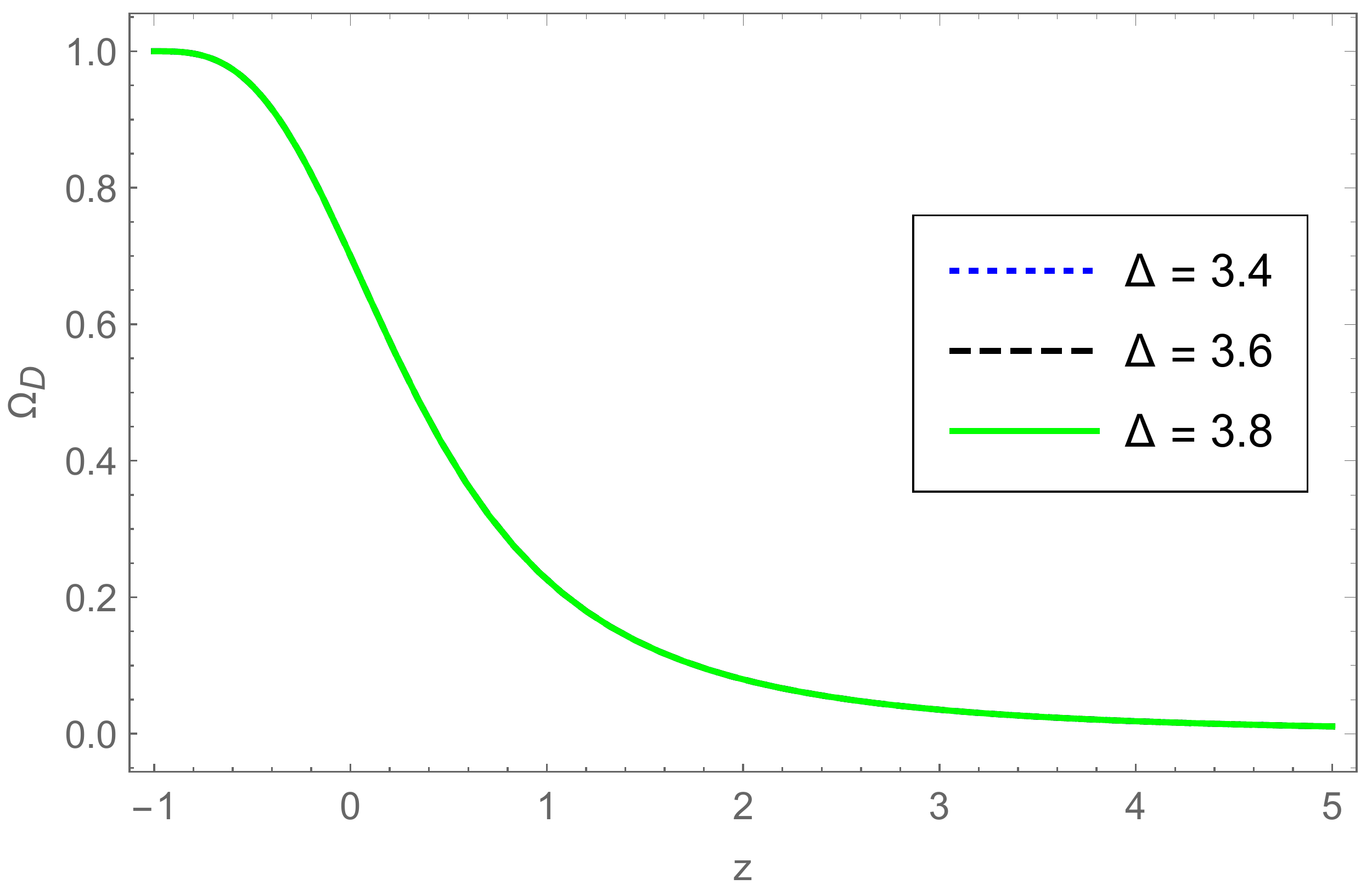}
	\caption{The evolutionary behaviour of Barrow  energy density parameter  $\Omega_{D}$ versus redshift for non-interacting BADE. Here, we have considered $H_{0}=67$, $\Omega_{D0} =0.70$ and $C=10$.} 
\end{figure}
\begin{eqnarray}
\label{eq13}
v^{2}_{s} = \frac{dp_{D}}{d\rho_{D}} = \frac{\dot{p_{D}}}{\dot\rho_{D}} = \frac{\rho_{D}}{\dot{\rho_{D}}} \dot{\omega_{D}} + \omega_{D},
\end{eqnarray}

at last we reaches to

\begin{eqnarray}
\label{eq14}
v^{2}_{s} = \frac{\Omega_{D}-3}{2} + \frac{3^{\frac{1}{2-\Delta}} (H^{2}\Omega_{D} C^{-1})^{\frac{1}{2-\Delta}} (6-2\Delta+(\Delta-2)\Omega_{D})}{6H}.
\end{eqnarray}

\begin{figure}[htbp]
	\centering
	\includegraphics[width=8cm,height=8cm,angle=0]{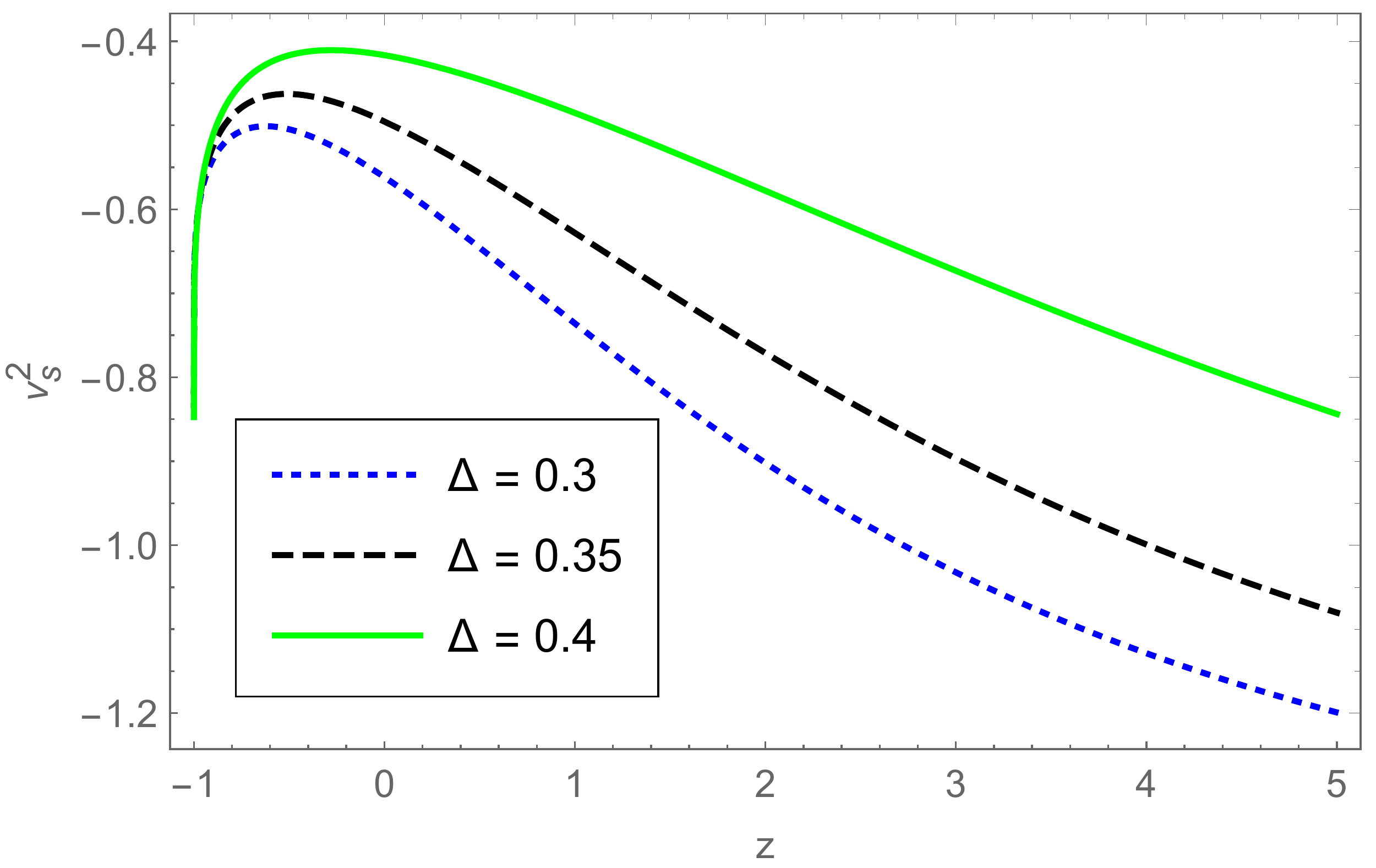}
	\includegraphics[width=8cm,height=8cm,angle=0]{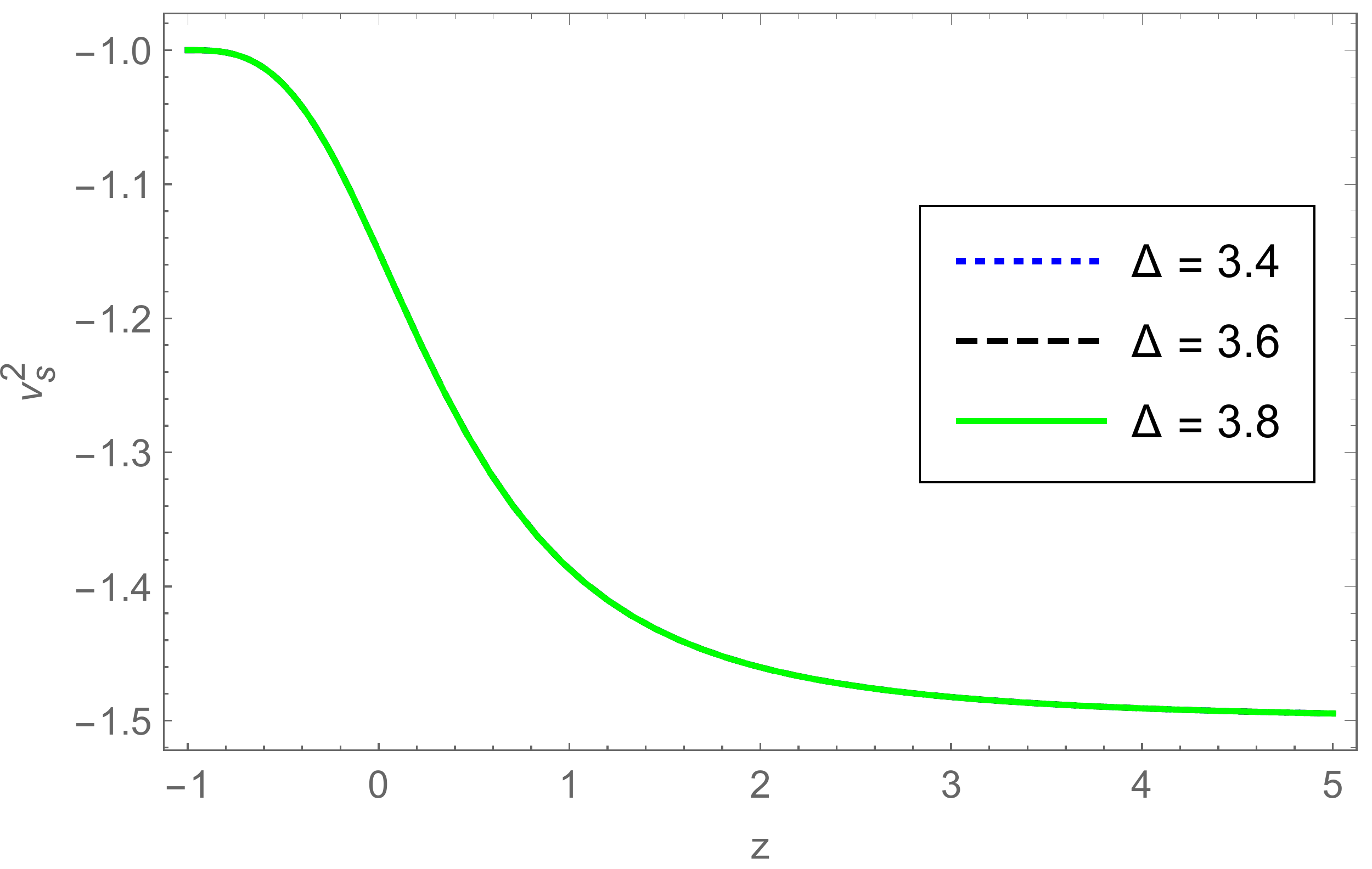}
	\caption{ The evolutionary behaviour of the squared sound speed $v^{2}_{s}$ versus redshift for non-interacting BADE. Here, we have considered $H_{0}=67$, $\Omega_{D0} =0.70$ and $C=10$.} 
\end{figure}
for the non-interacting case. In Figs. 1-4, we have plotted  the evolutionary behaviour of the system parameters. From the squared sound speed, it is obvious that the model is classically unstable ($v^{2}_{s} < 0$). Furthermore,  $\omega_{D}$, $q$ and $\Omega_{D}$ themselves show adequate behaviour for different values of $\Delta$. 


\subsection{Interacting case (Q $\neq$ 0)}

Current observations symbolize that the evolution of DE and DM is not self-governing as specified ahead and a key to explain the coincidence puzzle \cite{ref62}. To find the expressions for the equation of state parameter, DP,  density parameters, and $v^{2}_{s}$, we assumed $Q = 3 b^{2} (\rho_{D}+\rho_{m})$ mutual interaction between  the DM and DE.

\begin{eqnarray}
\label{eq15}
\omega_{D} = -1 -\frac{b^{2}}{\Omega_{D}} -  \frac{\Delta - 2}{3 H T}
\end{eqnarray}
\begin{figure}[htbp]
	\centering
	\includegraphics[width=8cm,height=8cm,angle=0]{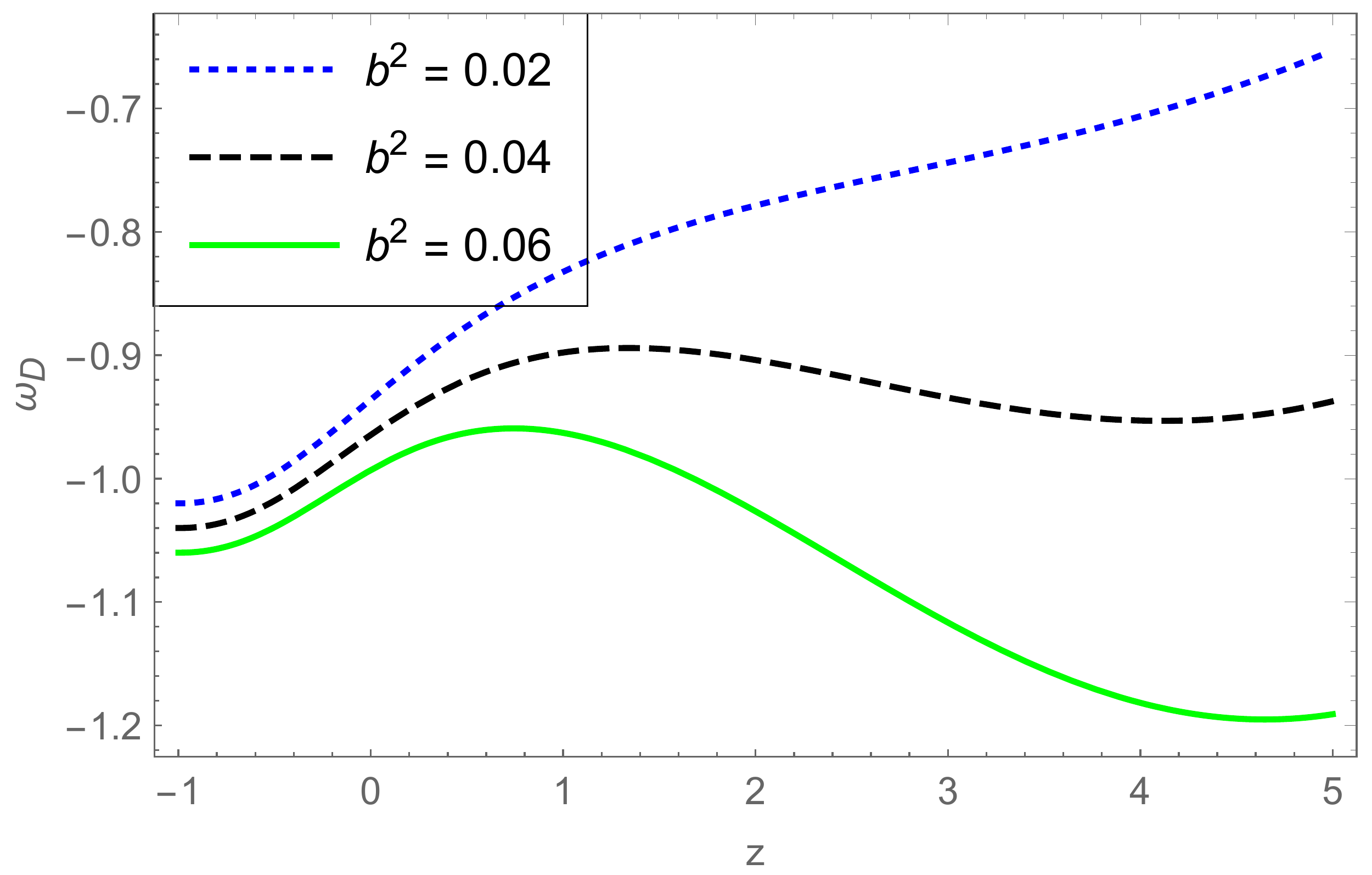}
	\caption{ The evolutionary behaviour of equation of state parameter  $\omega_{D}$ versus redshift for interacting BADE. Here, we have considered $H_{0}=67$, $\Omega_{D0} =0.70$, $\Delta = - 5$ and $C=10$.} 
\end{figure}
\begin{eqnarray}
\label{eq16}
\frac{\dot{H}}{H^{2}} =  \frac{3}{2} (-1+b^{2}+\Omega_{D}) + \frac{\Omega_{D} (\Delta - 2)}{2HT}
\end{eqnarray}

\begin{eqnarray}
\label{eq17}
q = \frac{1}{2} (1 - 3 b^{2} - 3\Omega_{D})  - \frac{\Omega_{D} (\Delta - 2)}{2HT}
\end{eqnarray}
\begin{figure}[htbp]
	\centering
	\includegraphics[width=8cm,height=8cm,angle=0]{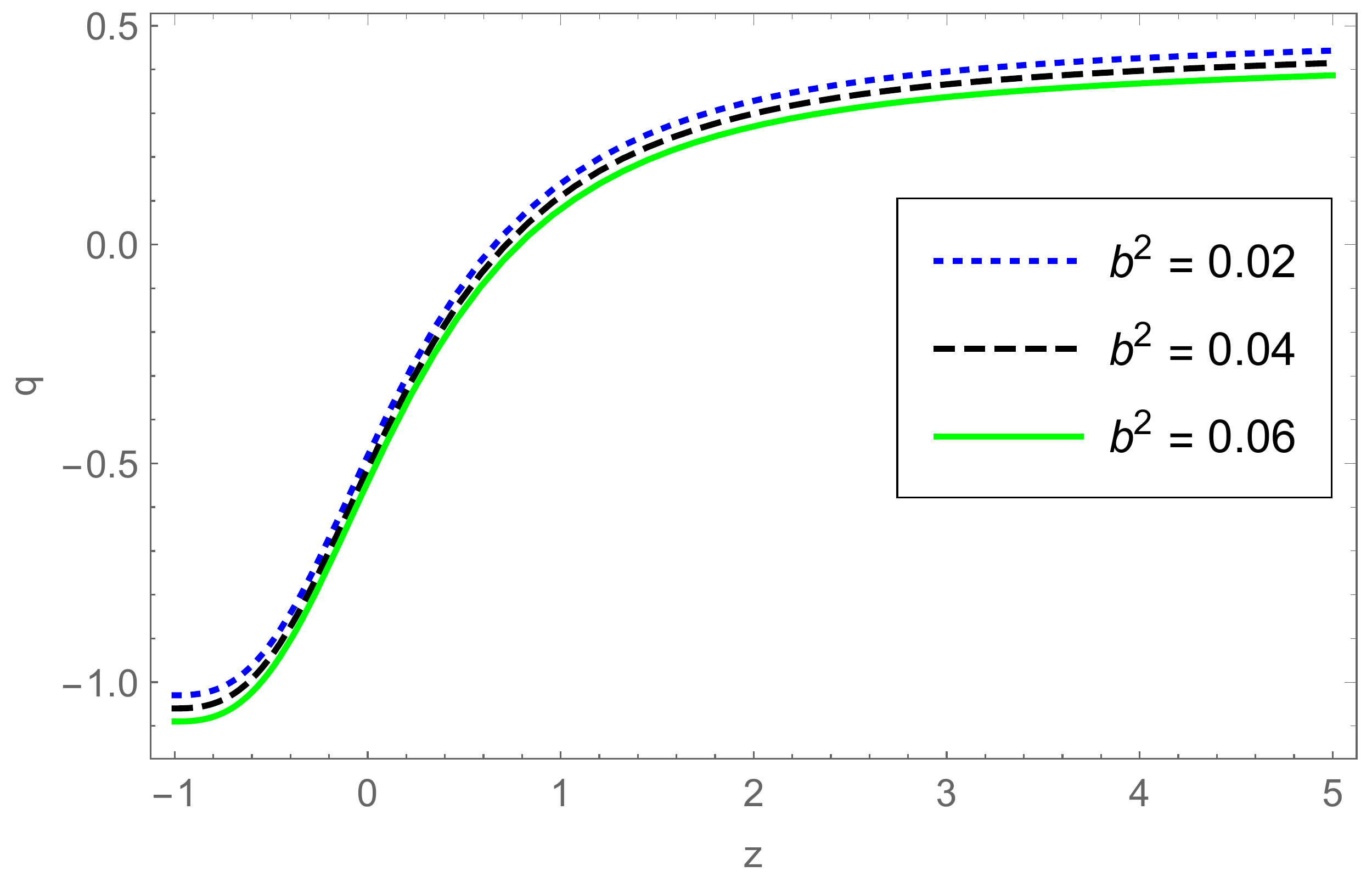}
	\caption{The evolutionary behaviour of the deceleration parameter  $q$ versus redshift for interacting BADE. Here, we have considered $H_{0}=67$, $\Omega_{D0} =0.70$, $\Delta = - 5$ and $C=10$.} 
\end{figure}
\begin{eqnarray}
\label{eq18}
 \dot{\Omega_{D}} = 2 H \Omega_{D}(1+q) + \frac{\Omega_{D} (\Delta - 2)}{T}
\end{eqnarray}
\begin{figure}[htbp]
	\centering
	\includegraphics[width=8cm,height=8cm,angle=0]{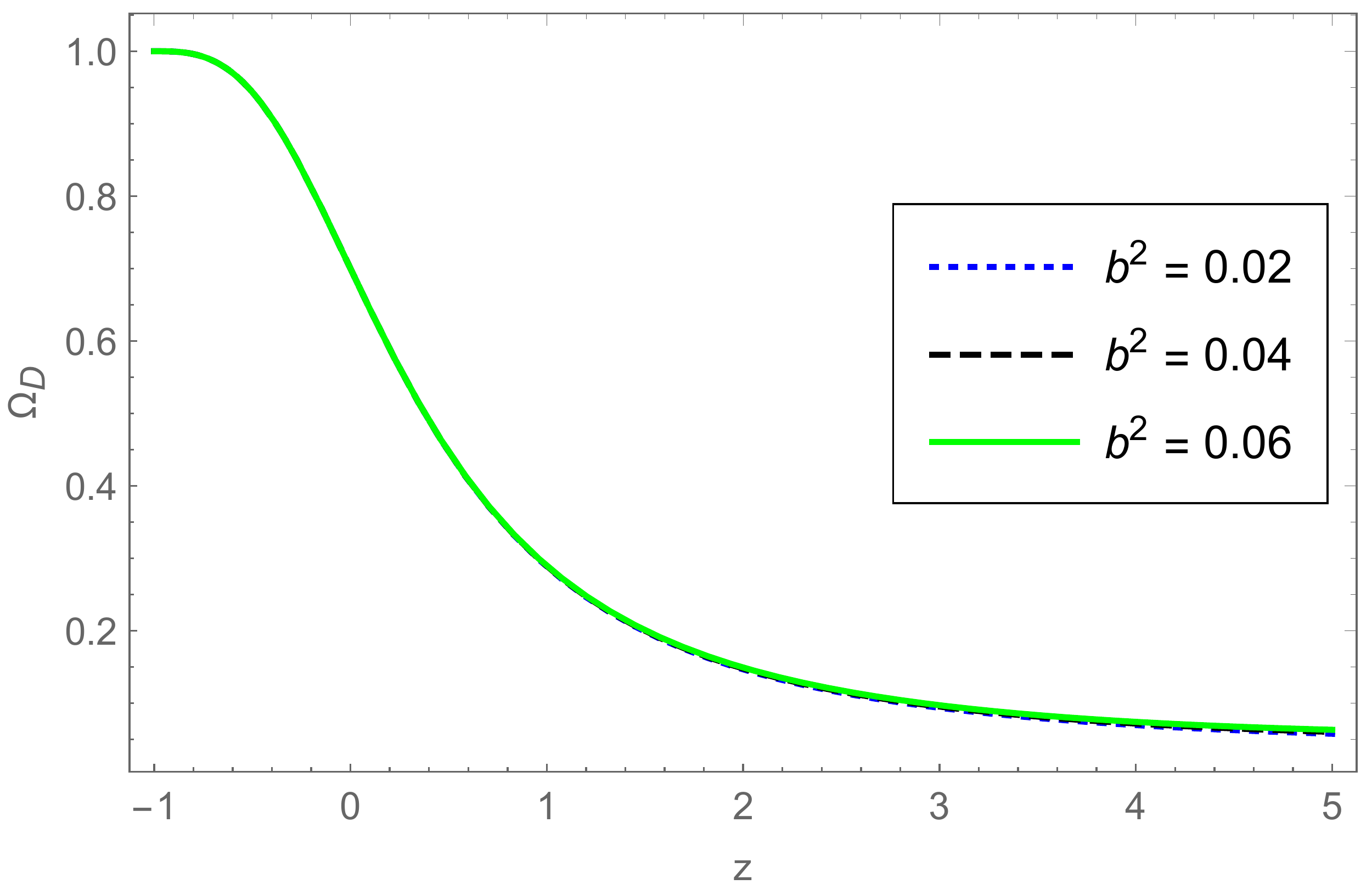}
	\caption{The evolutionary behaviour of the Barrow energy density parameter $\Omega_{D}$ versus redshift for interacting BADE. Here, we have considered $H_{0}=67$, $\Omega_{D0} =0.70$, $\Delta = - 5$ and $C=10$.} 
\end{figure}
\begin{eqnarray}
\label{eq19}
v^{2}_{s} = \frac{-3 - b^{2} + \Omega_{D}}{2} - \frac{3^{\frac{\Delta-3}{\Delta-2}} b^{2} H (H^{2}\Omega_{D} C^{-1})^{\frac{1}{\Delta - 2}} (-1 + b^{2}+\Omega_{D})}{\Omega_{D}(\Delta-2)} \nonumber\\
+ \frac{3^{\frac{1}{-\Delta+2}} (H^{2}\Omega_{D} C^{-1})^{\frac{1}{-\Delta+2}} (6-2\Delta+(\Delta-2)\Omega_{D})}{6H}
\end{eqnarray}

\begin{figure}[htbp]
	\centering
	\includegraphics[width=8cm,height=8cm,angle=0]{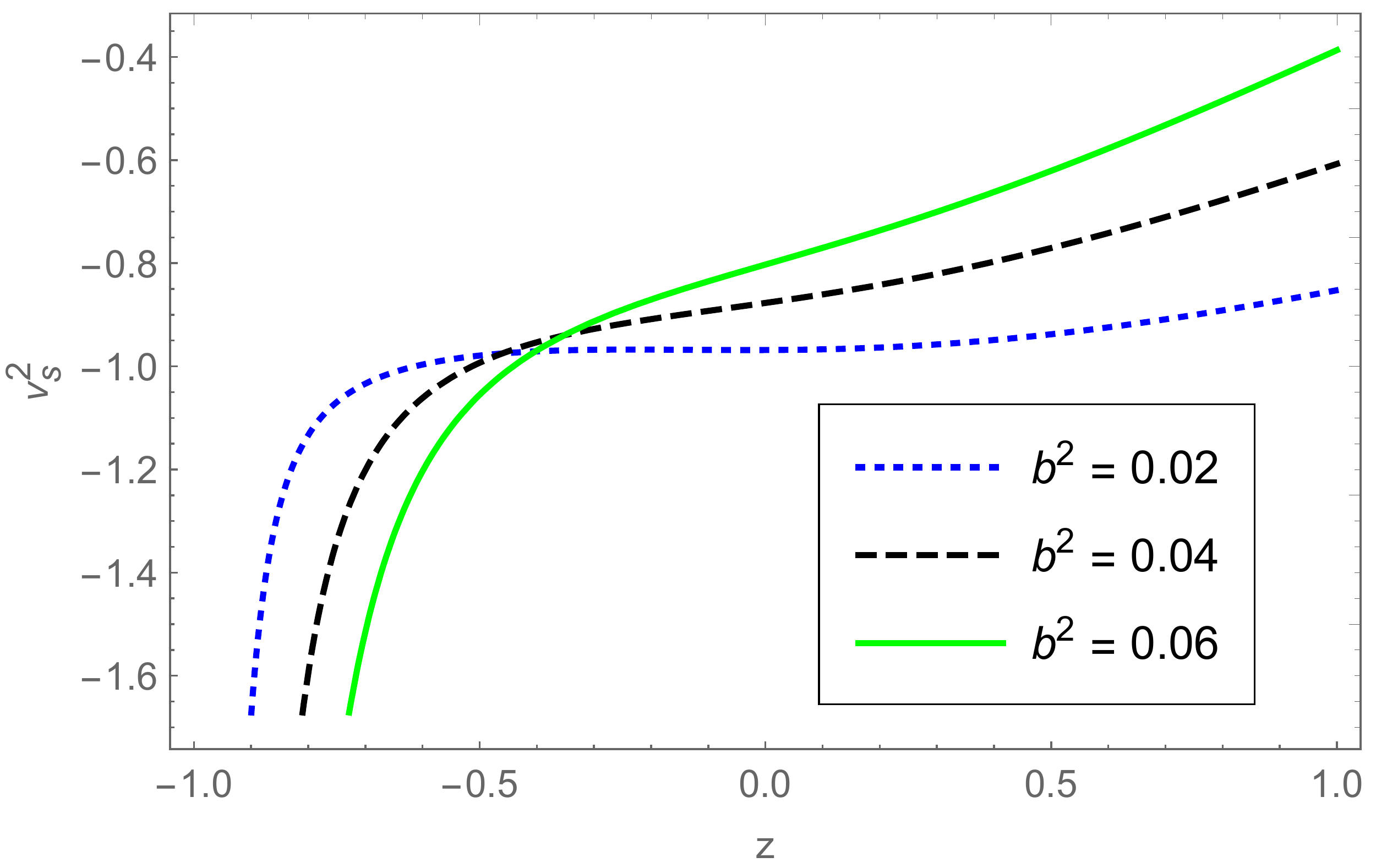}
	\caption{ The evolutionary behaviour the squared sound speed  $v_{s}^{2}$ versus redshift for interacting BADE. Here, we have considered $H_{0}=67$, $\Omega_{D0} =0.70$, $\Delta = - 5$ and $C=10$.} 
\end{figure}
The evolutionary behaviour  of system parameters ($\omega_{D}$, $q$ and  $\Omega_{D}$) is plotted in Figs. 5-8 for interacting BADE model which present satisfactory behaviours. Also, it is clear that this case is classically unstable,  unlike the nature of Tsallis agegraphic dark energy (TADE) \cite{ref63}.

\section{New Barrow agegraphic dark energy (NBADE) model }

Applying  $\eta$ (the conformal time) as the infrared cut-off rather than the  universe age, Wei and Cai \cite{ref64} introduced a new agegraphic DE due to some issues in the original ADE \cite{ref21}. Here, $\eta$ is interpreted as $ a d\eta=dt$ heading to $\dot{\eta} = 1/a$ and next

\begin{eqnarray}
\label{eq20}
\eta = \int_{0}^{a} \frac{da}{ a^{2} H}.
\end{eqnarray}

In this way, The energy density of NBADE can be formulate using  Eq. (\ref{eq3}) as

\begin{eqnarray}
\label{eq21}
\rho_{D} = C \eta^{\Delta-2}
\end{eqnarray}

\subsection{Non-interacting case}
 We can include Eq. (\ref{eq21}) and its derivative with time into Eq. (\ref{eq4b}), as there is no interaction present between the DM and DE $(Q = 0)$ to enter
 
\begin{eqnarray}
\label{eq22}
\omega_{D} = -1 -  \frac{\Delta - 2}{3 a H \eta}
\end{eqnarray}

here $\eta$ presents $\left(\frac{3 H^{2} \Omega_{D}}{C}\right)^{\frac{1}{\Delta-2}}$. Differentiating Eq.(\ref{eq5}) and practicing Eqs. (\ref{eq22}) and (\ref{eq4a}), we get

\begin{figure}[htbp]
	\centering
	\includegraphics[width=8cm,height=8cm,angle=0]{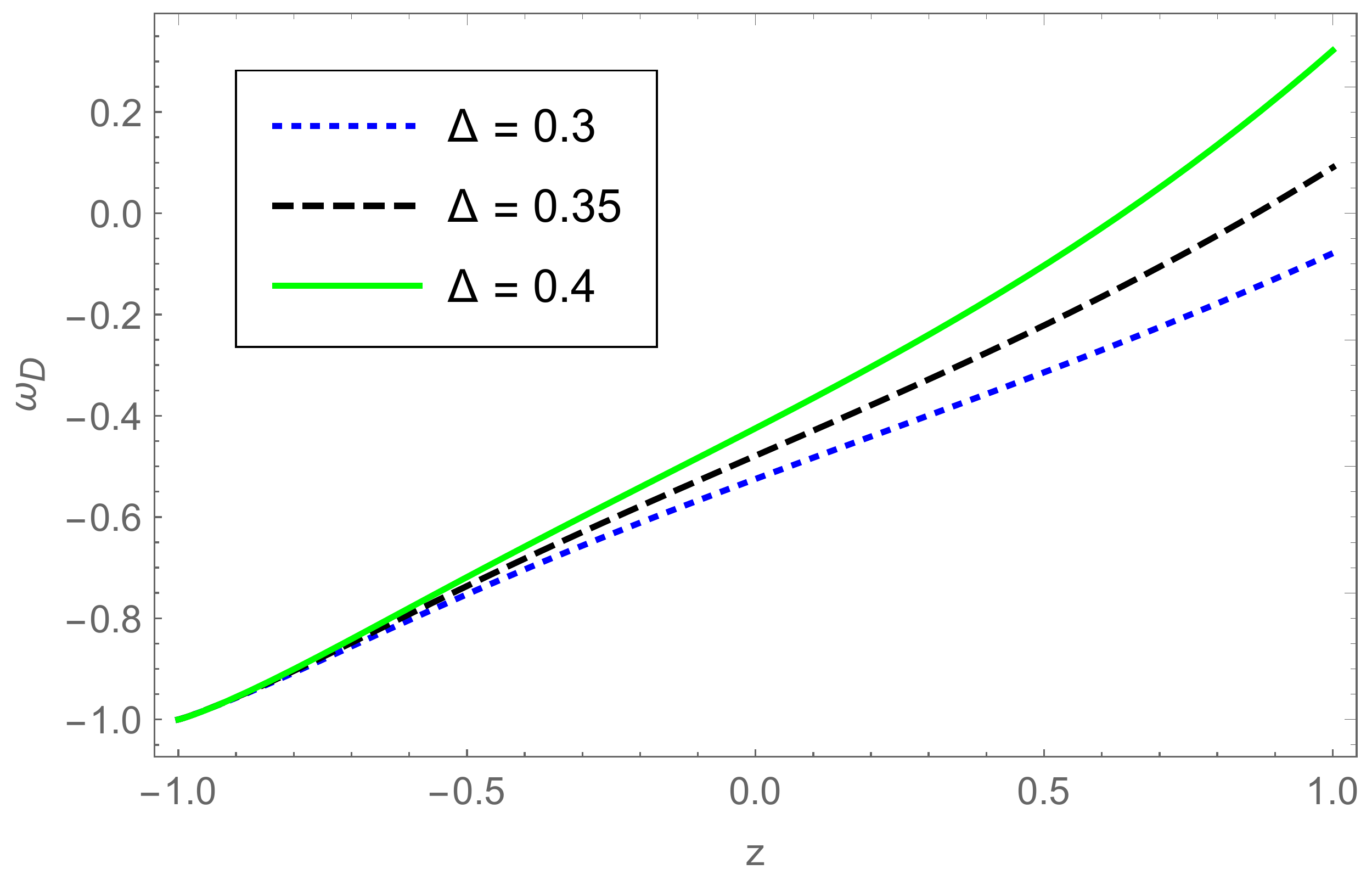}
	\includegraphics[width=8cm,height=8cm,angle=0]{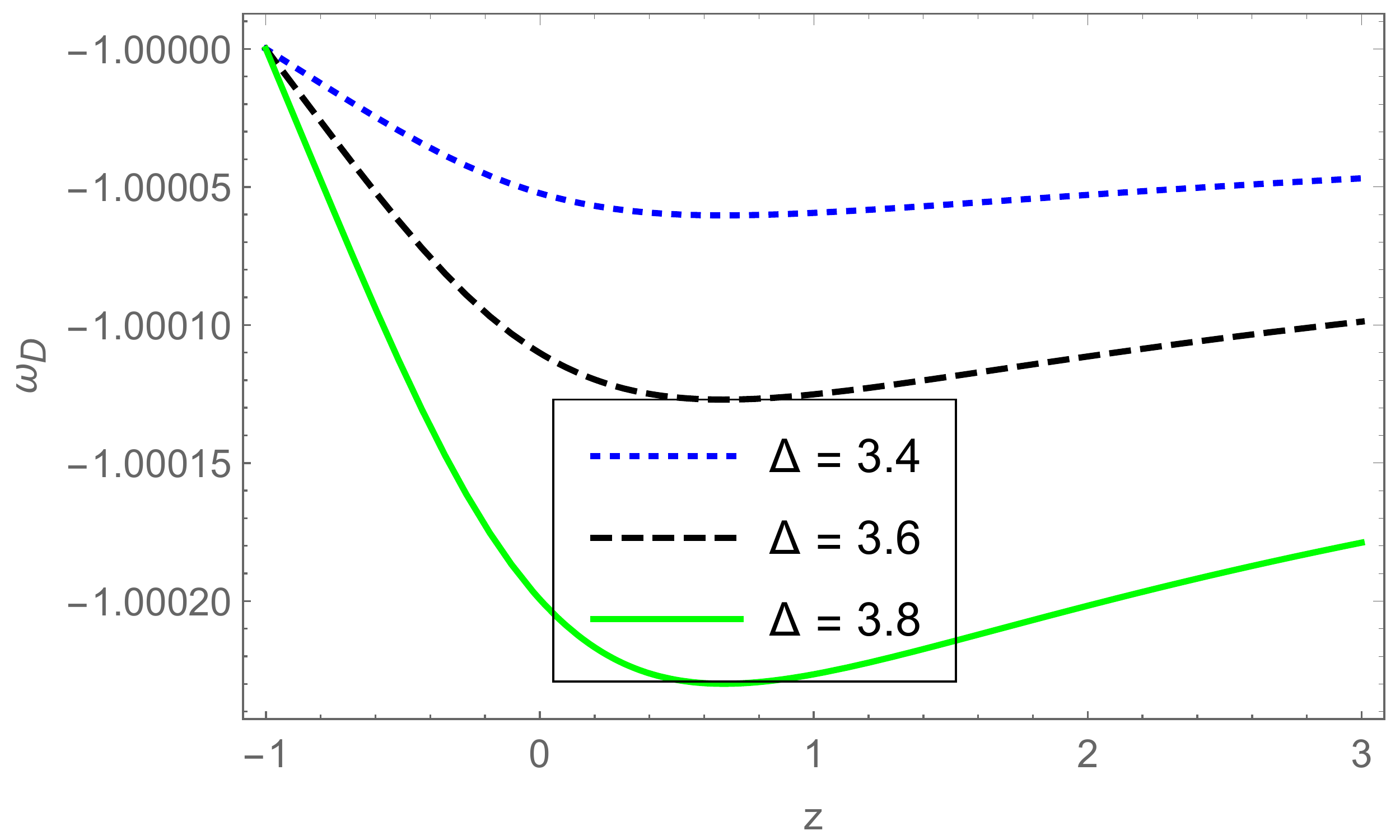}
	\caption{The evolutionary behaviour of equation of state parameter  $\omega_{D}$ versus redshift for non-interacting NBADE. Here, we have considered $H_{0}=67$, $\Omega_{D0} =0.70$ and $C=10$.} 
\end{figure}
\begin{eqnarray}
\label{eq23 }
\frac{\dot{H}}{H^{2}} = - \frac{3}{2} (1-\Omega_{D}) + \frac{\Omega_{D} (\Delta - 2)}{2 a H \eta}
\end{eqnarray}

\begin{eqnarray}
\label{eq24}
q = \frac{1}{2} (1-3\Omega_{D})  - \frac{\Omega_{D} (\Delta - 2)}{2 a H \eta}
\end{eqnarray}
\begin{figure}[htbp]
	\centering
	\includegraphics[width=8cm,height=8cm,angle=0]{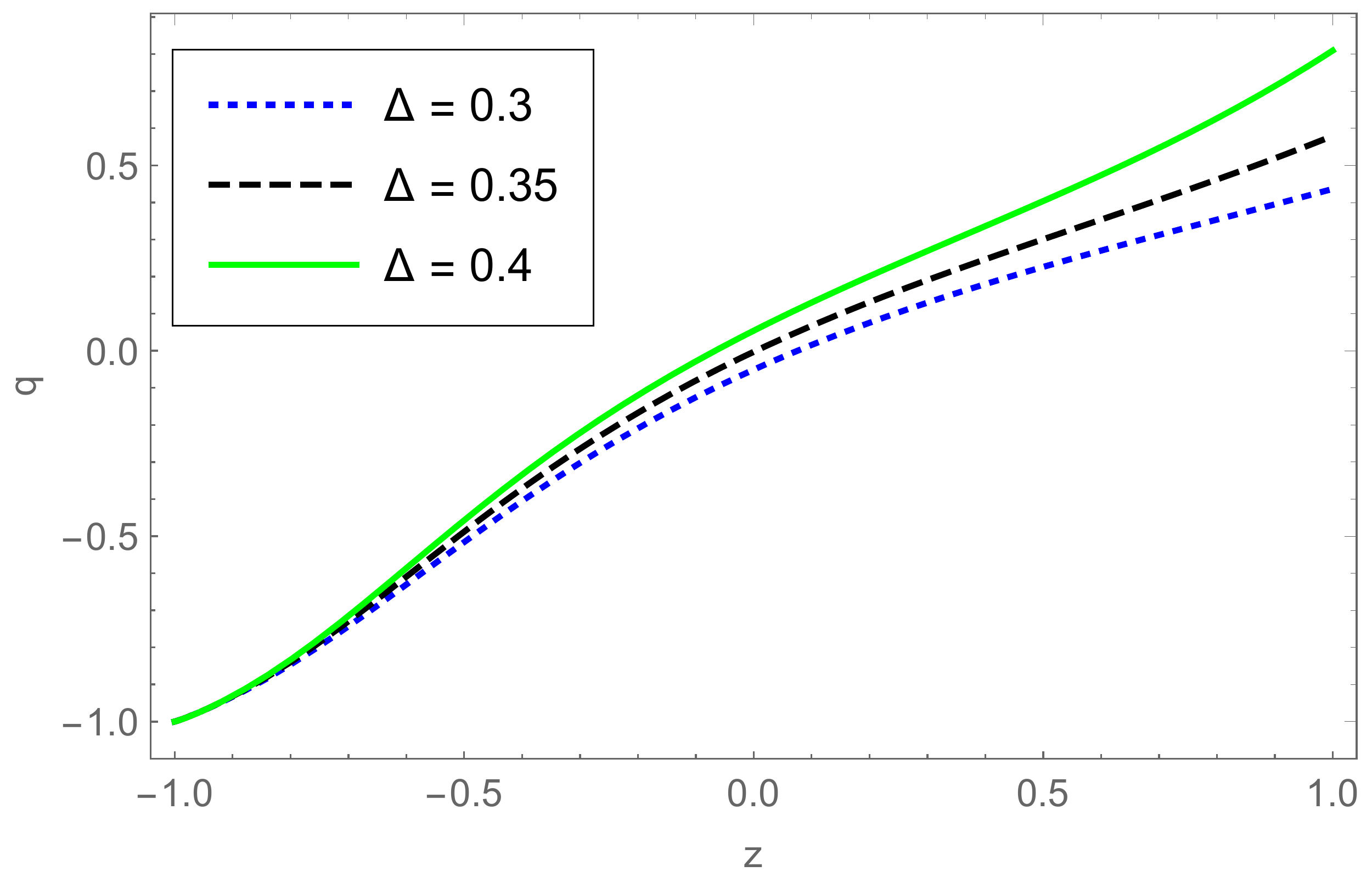}
	\includegraphics[width=8cm,height=8cm,angle=0]{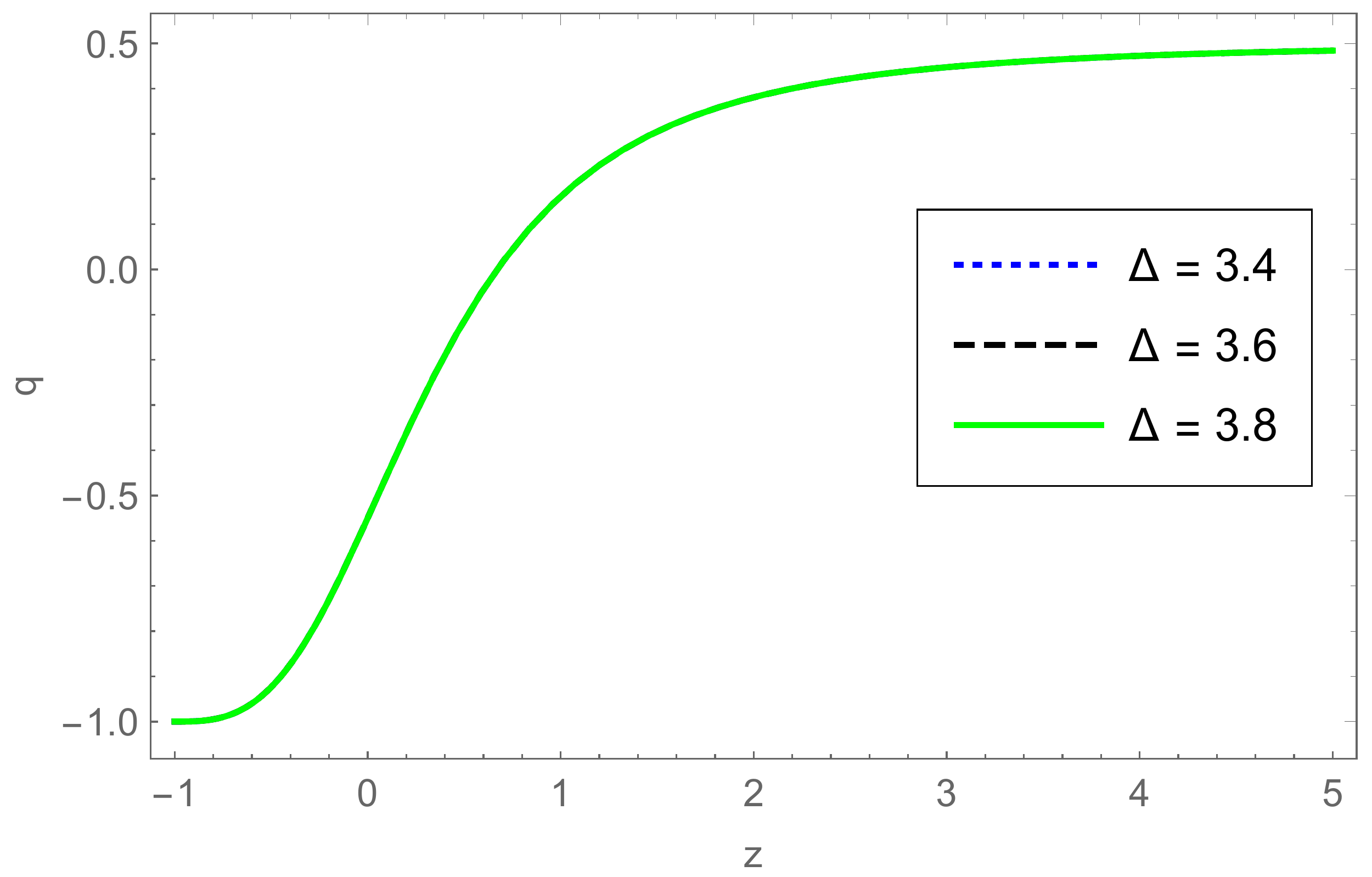}
	\caption{The evolutionary behaviour of the deceleration parameter  $q$ versus redshift for non-interacting NBADE. Here, we have considered $H_{0}=67$, $\Omega_{D0} =0.70$ and $C=10$.} 
\end{figure}
shows the deceleration parameter. Additionally, it is a matter of estimation to apply Eqs. (\ref{eq7}) and (\ref{eq22}) to confirm that

\begin{eqnarray}
\label{eq25}
\dot{\Omega_{D}} =  2 H \Omega_{D}(1+q) + \frac{\Omega_{D} (\Delta - 2)}{ a \eta} 
\end{eqnarray}
\begin{figure}[htbp]
	\centering
	\includegraphics[width=8cm,height=8cm,angle=0]{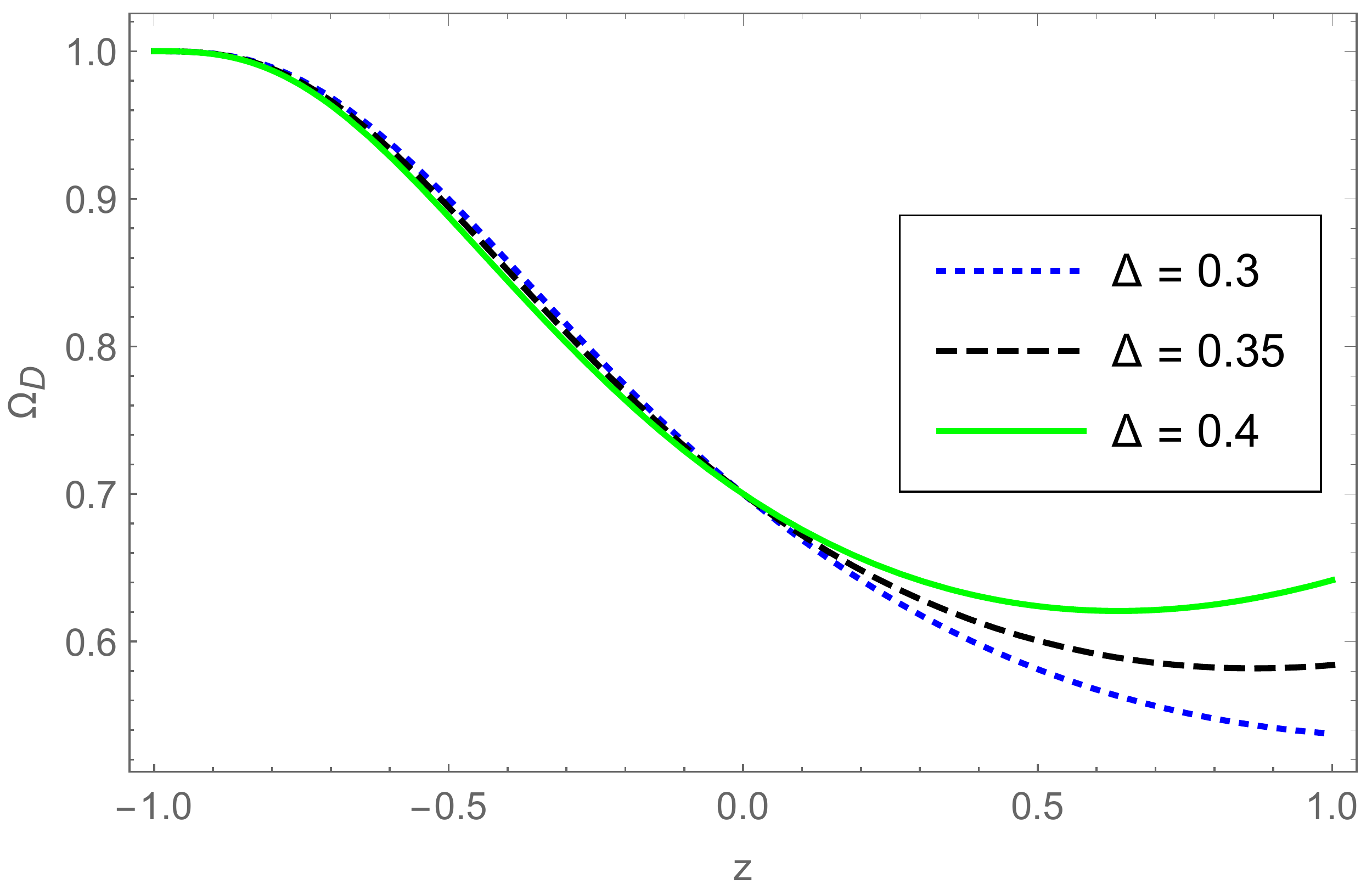}
	\includegraphics[width=8cm,height=8cm,angle=0]{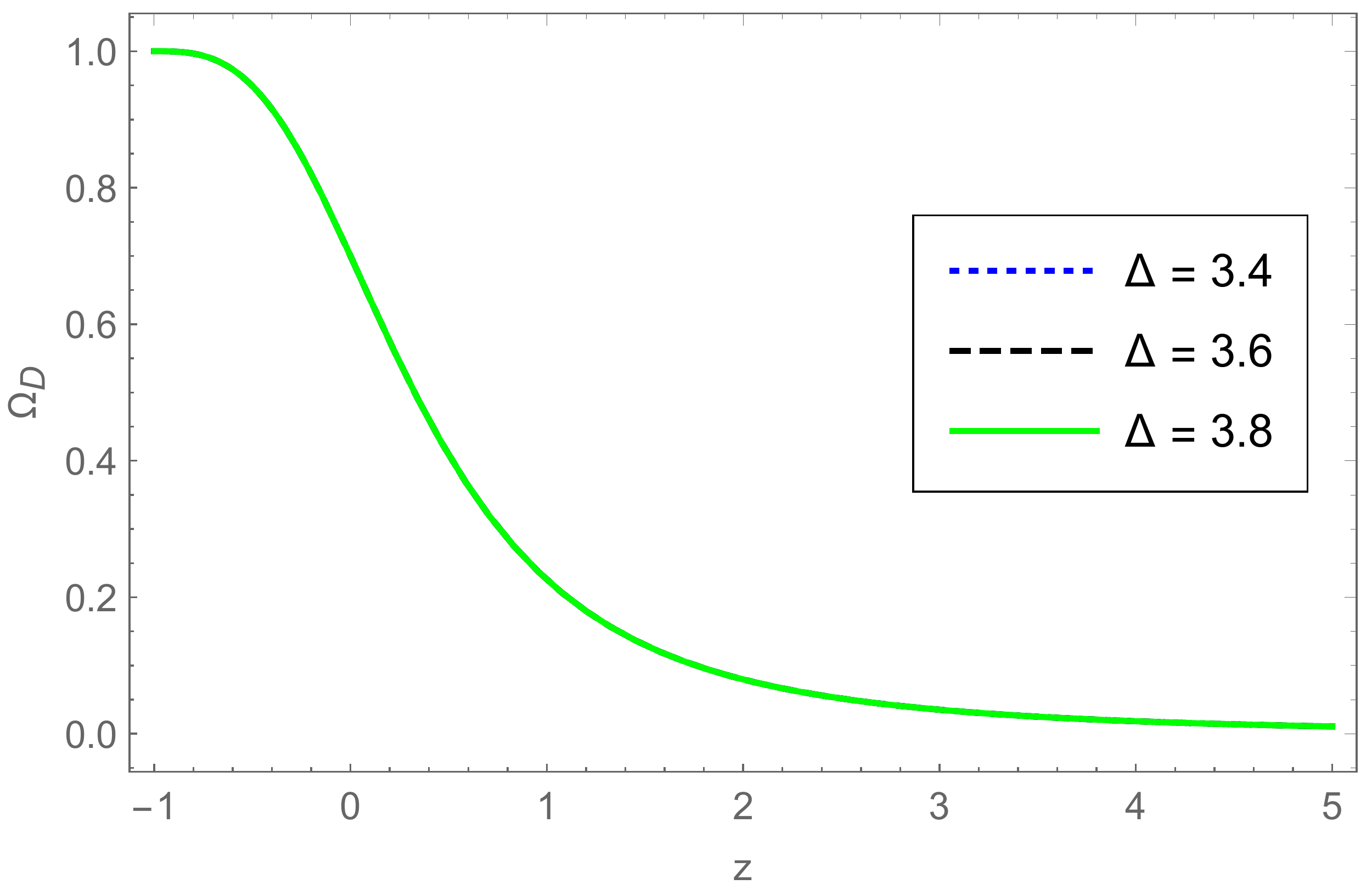}
	\caption{The evolutionary behaviour of the Barrow energy density parameter  $\Omega_{D}$ versus redshift for non-interacting NBADE. Here, we have considered $H_{0}=67$, $\Omega_{D0} =0.70$ and $C=10$.} 
\end{figure}
Lastly, we get the squared of the sound speed as

\begin{eqnarray}
\label{eq26}
v^{2}_{s} = \frac{-7+3\Omega_{D} }{6}
+ \frac{3^{\frac{1}{-\Delta+2}} (H^{2}\Omega_{D} C^{-1})^{\frac{1}{-\Delta+2}} (6-2\Delta+(\Delta-2)\Omega_{D})}{6 a H}
\end{eqnarray}
\begin{figure}[htbp]
	\centering
	\includegraphics[width=8cm,height=8cm,angle=0]{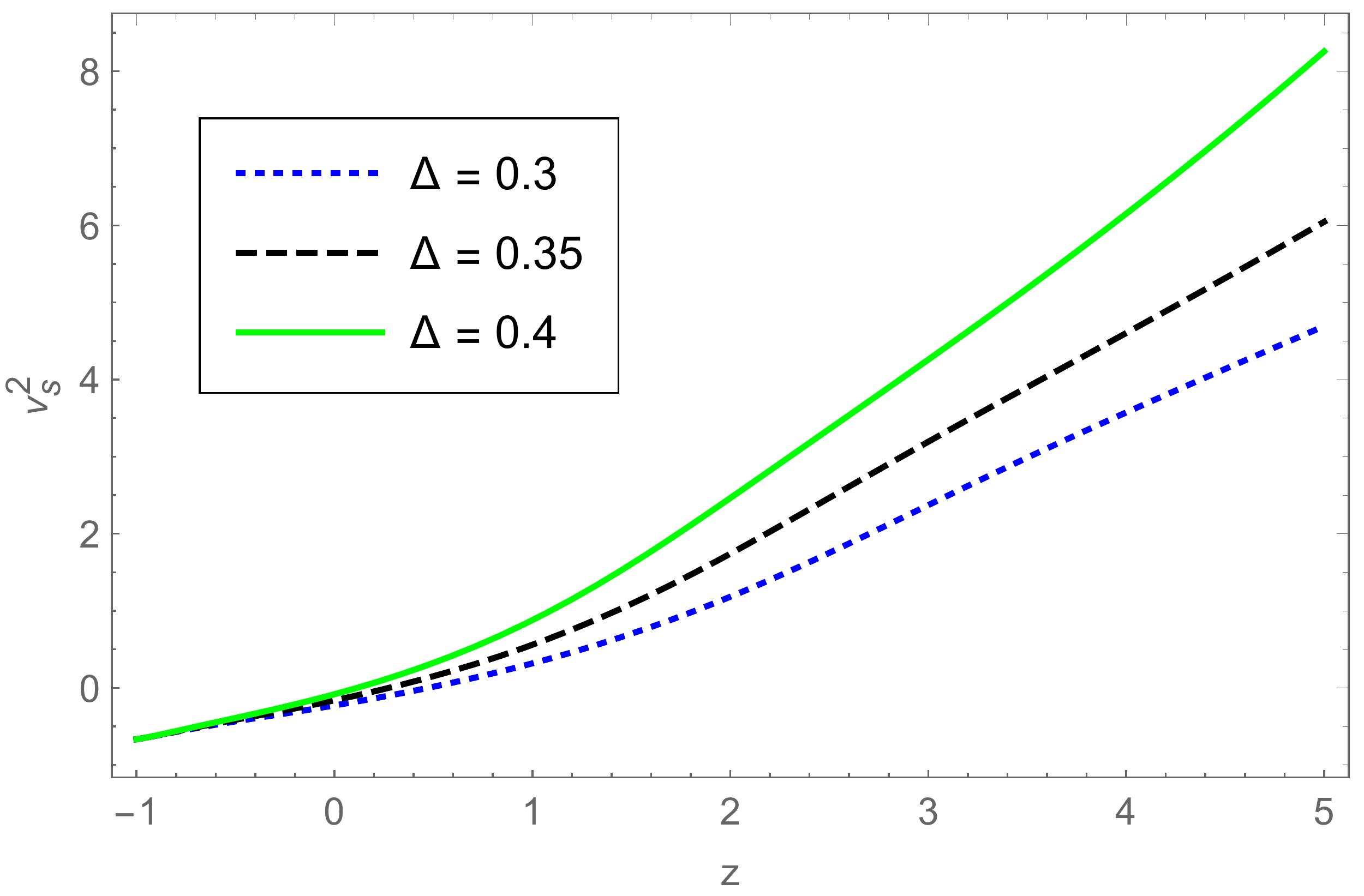}
	\includegraphics[width=8cm,height=8cm,angle=0]{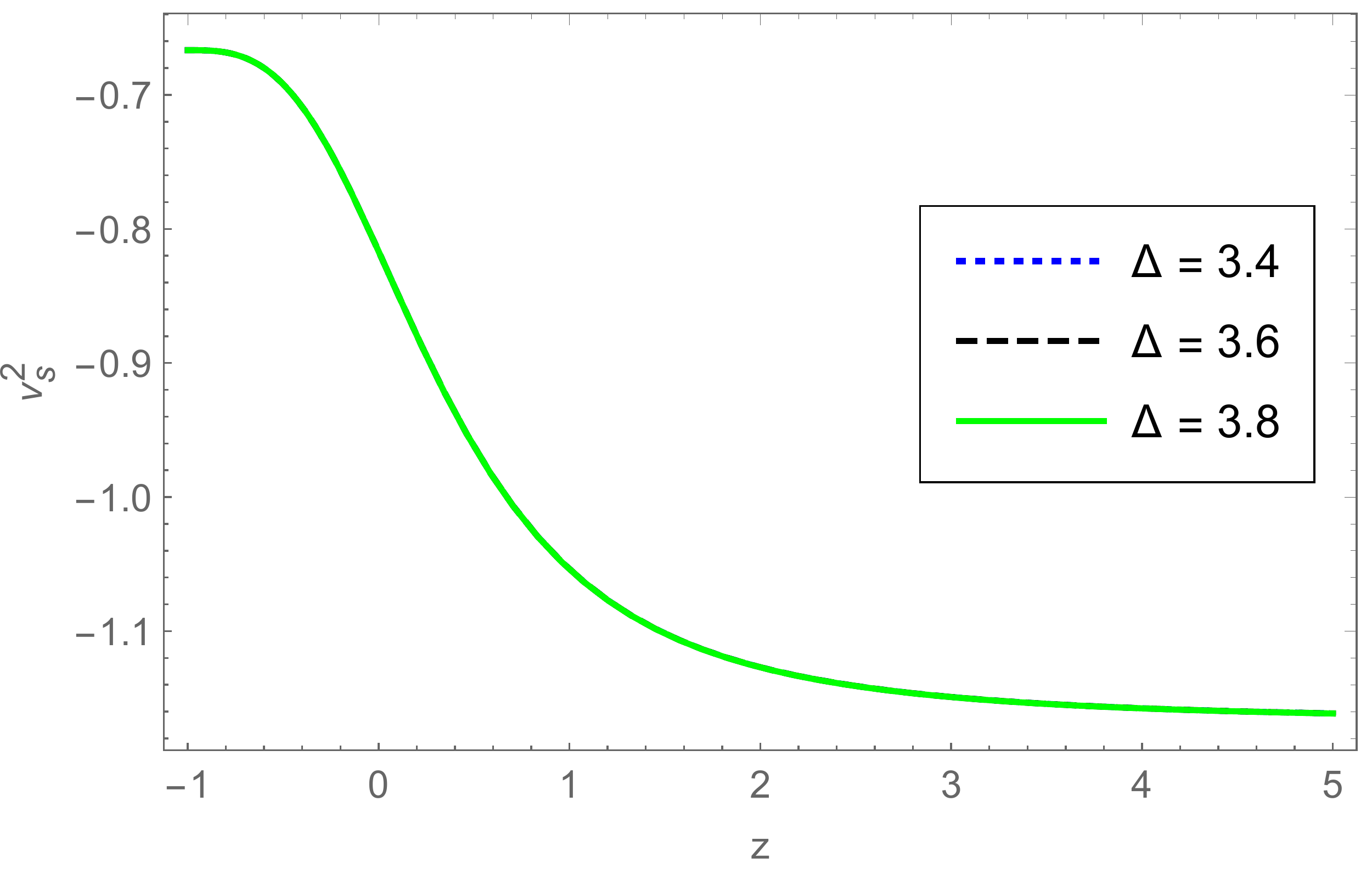}
	\caption{The evolutionary behaviour of the squared sound speed  $v_{s}^{2}$ versus redshift for non-interacting NBADE. Here, we have considered $H_{0}=67$, $\Omega_{D0} =0.70$ and $C=10$.} 
\end{figure}

Figs. 9-12 depicts the evolution of the system parameters for non-interacting NBADE declaring that, even though $q$, $\Omega_{D}$, and $\omega_{D}$ can themselves present the satisfactory nature throughout the cosmic evolution, and the model is stable ($v^{2}_{s}\geq0$) in the past and present for $\Delta= 0.3, 0.35$ and 0.4, while the model is unstable ($v^{2}_{s}<0$)  for $\Delta= 3.4, 3.6$ and 3.8.

\begin{figure}[htbp]
	\centering
	\includegraphics[width=8cm,height=8cm,angle=0]{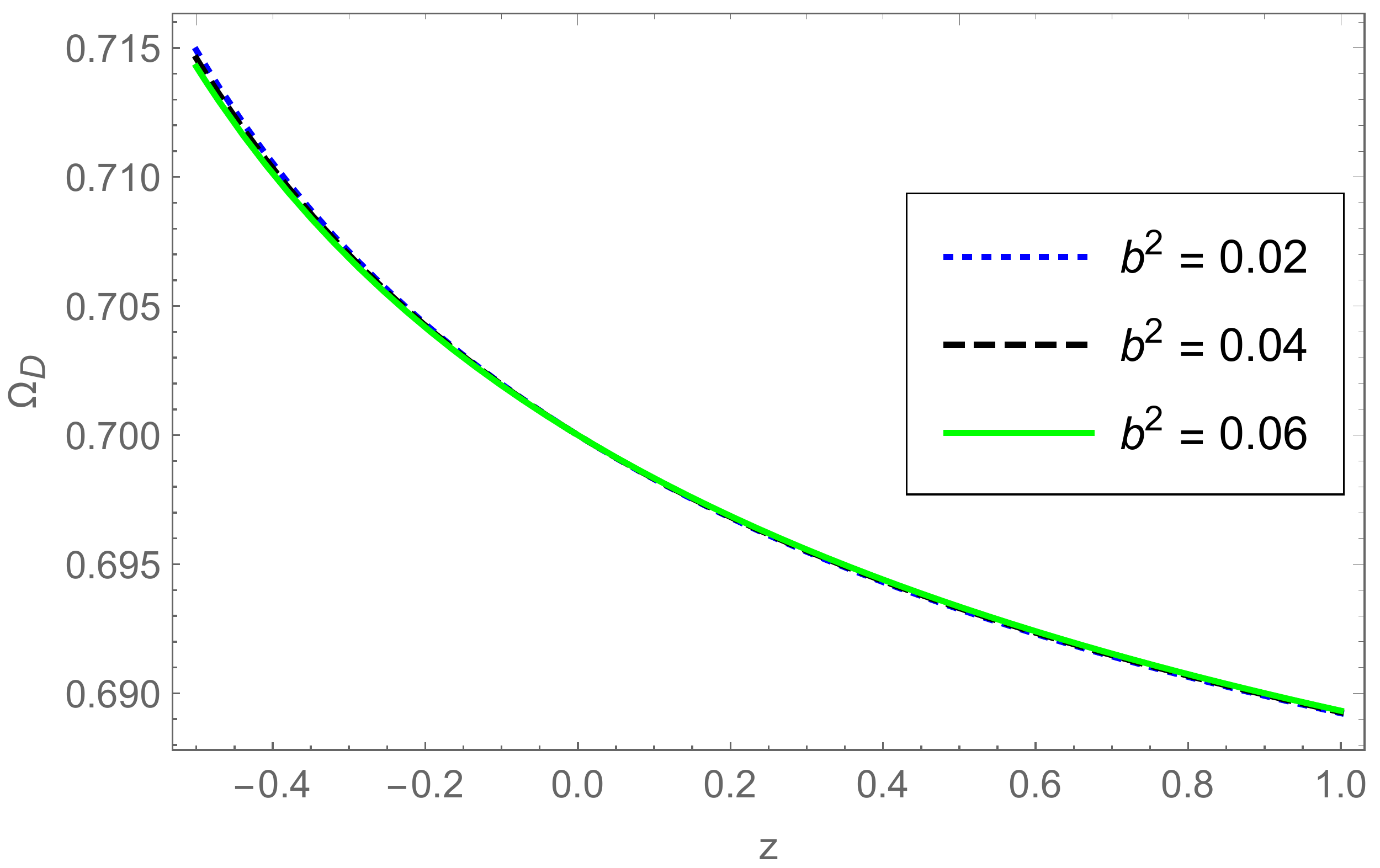}
	\caption{The evolutionary behaviour of the Barrow energy density parameter  $\Omega_{D}$ versus redshift for interacting NBADE. Here, we have considered $H_{0}=67$, $\Omega_{D0} =0.70$, $\Delta = - 5$ and $C=10$.} 
\end{figure}

\subsection{Interacting case}

It is a focus of calculations to discover the following system parameters, acknowledging the $Q=3b^{2}H(\rho_{D}+\rho_{m})$  interaction between the DM and DE of the universe.

\begin{eqnarray}
\label{eq27}
\omega_{D} = -1 -\frac{b^{2}}{\Omega_{D}} -  \frac{\Delta - 2}{3 a H \eta}
\end{eqnarray}
\begin{figure}[htbp]
	\centering
	\includegraphics[width=8cm,height=8cm,angle=0]{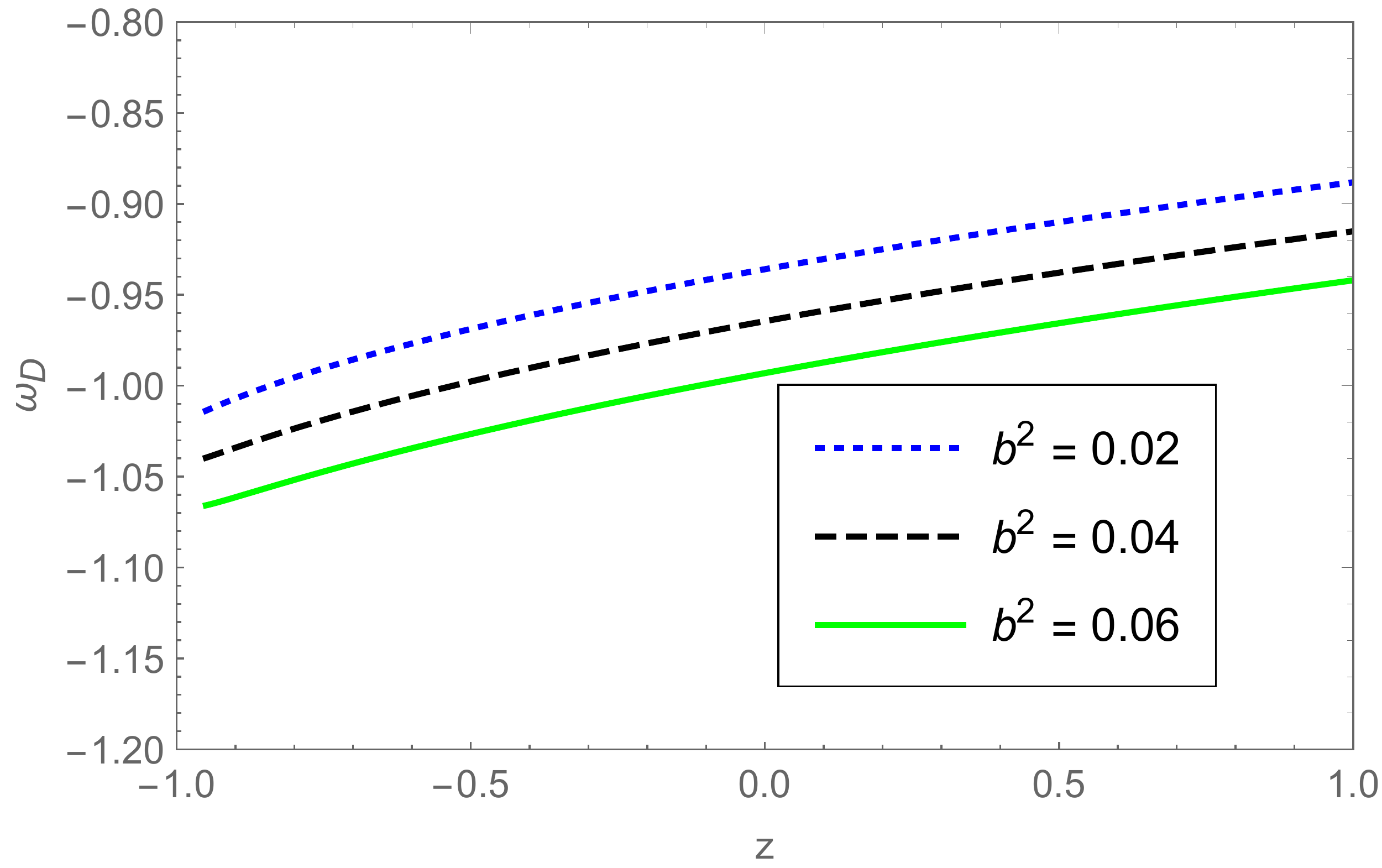}
	\caption{The evolutionary behaviour of equation of state parameter  $\omega_{D}$ versus redshift for interacting NBADE. Here, we have considered $H_{0}=67$, $\Omega_{D0} =0.70$, $\Delta = - 5$ and $C=10$.} 
	\end{figure}
\begin{eqnarray}
\label{eq28}
\frac{\dot{H}}{H^{2}} =  \frac{3}{2} (-1+b^{2}+\Omega_{D}) + \frac{\Omega_{D} (\Delta - 2)}{2 a H \eta}
\end{eqnarray}

\begin{eqnarray}
\label{eq29}
q = \frac{1}{2} (1 -  3 b^{2} - 3\Omega_{D})  - \frac{\Omega_{D} (\Delta - 2)}{2 a H \eta}
\end{eqnarray}
\begin{figure}[htbp]
	\centering
	\includegraphics[width=8cm,height=8cm,angle=0]{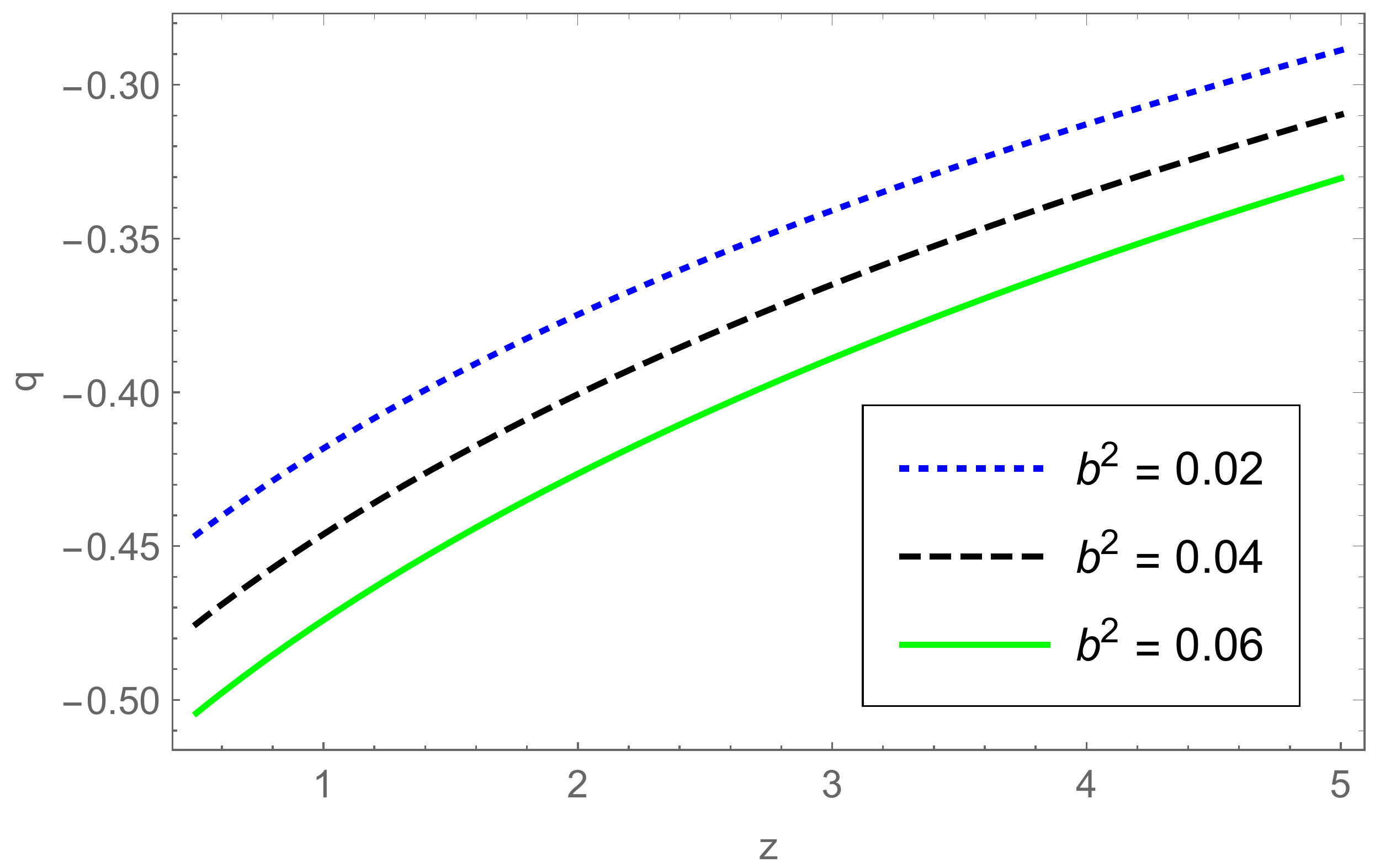}
	\caption{The evolutionary behaviour of the deceleration parameter  $q$ versus redshift for interacting NBADE. Here, we have considered $H_{0}=67$, $\Omega_{D0} =0.70$, $\Delta = - 5$ and $C=10$.} 
\end{figure}
\begin{eqnarray}
\label{eq30}
\dot{\Omega_{D}} = 2 H \Omega_{D}(1+q) + \frac{\Omega_{D} (\Delta - 2)}{a \eta}
\end{eqnarray}

\begin{eqnarray}
\label{eq31}
v^{2}_{s} = \frac{-7 - 3b^{2} + 3\Omega_{D}}{6} - \frac{3^{\frac{\Delta-3}{\Delta-2}} b^{2} a H (H^{2}\Omega_{D} C^{-1})^{\frac{1}{\Delta - 2}} (-1 + b^{2}+\Omega_{D})}{\Omega_{D}(\Delta-2)} \nonumber\\
+ \frac{3^{\frac{1}{-\Delta+2}} (H^{2}\Omega_{D} C^{-1})^{\frac{1}{-\Delta+2}} (6-2\Delta+(\Delta-2)\Omega_{D})}{6 a H}
\end{eqnarray}
\begin{figure}[htbp]
	\centering
	\includegraphics[width=8cm,height=8cm,angle=0]{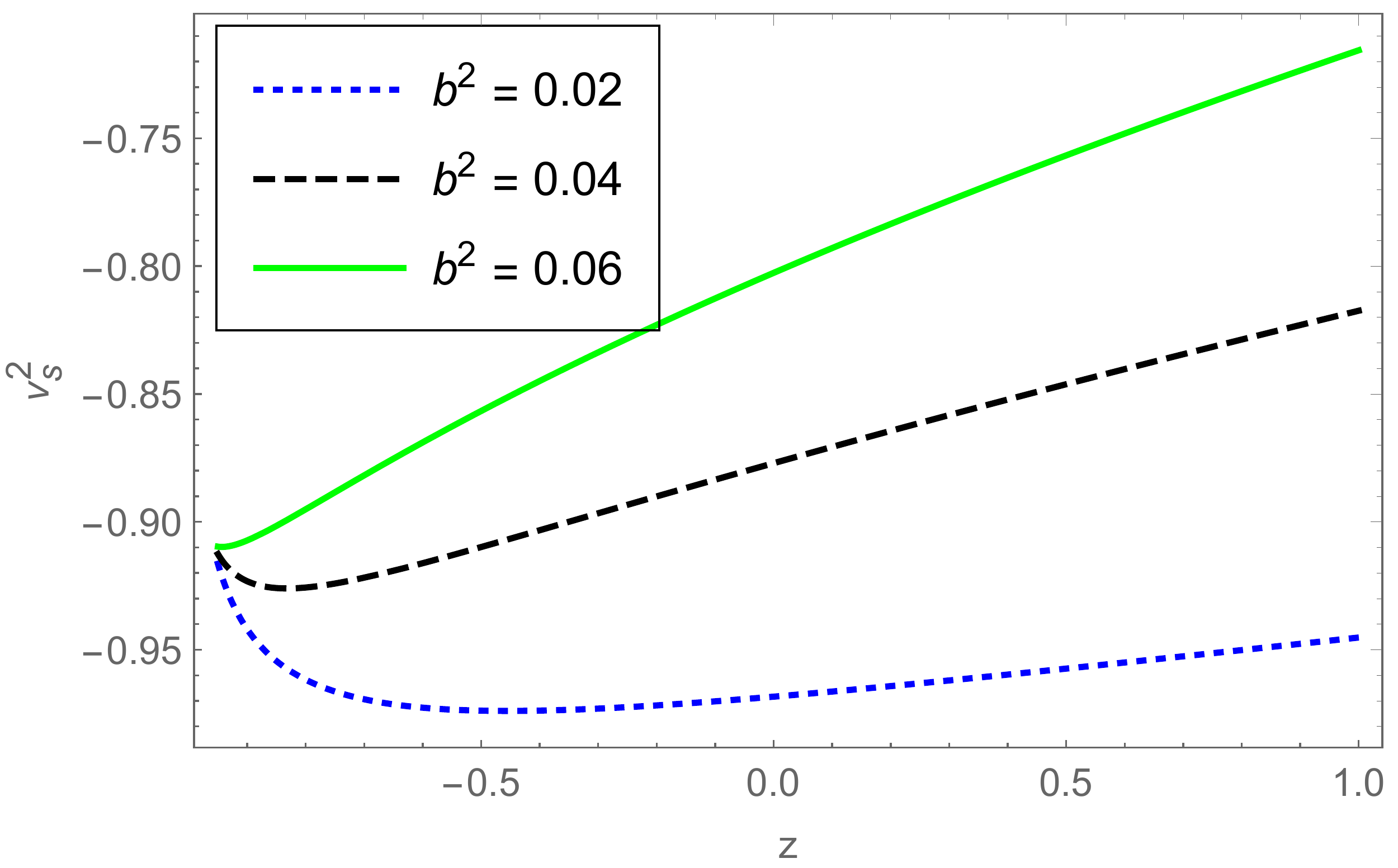}
	\caption{The evolutionary behaviour of the squared sound speed $v^{2}_{s}$ versus redshift for interacting NBADE. Here, we have considered $H_{0}=67$, $\Omega_{D0} =0.70$, $\Delta = - 5$ and $C=10$.} 
\end{figure}

These system parameters are plotted in Figs. 13-16. The parameters $\Omega_{D}$, $q$, and $\omega_{D}$ present a satisfactory nature, and the model is classically unstable. Identical to the BADE model, this model is also unstable at the standard level by the presence of $Q=3b^{2}H(\rho_{D}+\rho_{m})$ interaction among the cosmos sectors.\\

 We close the above discussion so far by mentioning the advantages of Barrow agegraphic dark energy over other dark energy models. The scenario of Barrow agegraphic dark energy has a new Barrow exponent $\Delta$. 
	It is important to mention here that  for the special case $\Delta = 0$, the Eq. (\ref{eq4})
	 provides the original agegraphic dark energy.
	Hence, the BADE is indeed a more general framework than the usual ADE scenario and therefore, we
	focused on the general case ($\Delta \neq 0$), where the quantum
	deformation effects switch on. We examined the role of $\Delta$ on the
	cosmological evolution. We extracted the simple differential equations for both BADE and NBADE with and without interaction for the evolution of the dark energy density parameter, and we presented the solution for the evolution of the corresponding
	dark energy equation-of-state parameter. As we showed,
	the scenario of Barrow agegraphic dark energy can describe the universe thermal history, with the sequence
	of matter and dark energy eras. Additionally, the new
	Barrow exponent $\Delta$ significantly affects the dark energy
	equation of state, and according to its value it can lead it
	to lie in the quintessence region, in the phantom region,
	or experience the phantom-divide crossing during the
	evolution. This was proved to be adequate
	for a successful description in agreement with observations, which serves as a significant advantage comparing
	to original ADE. 
	 Therfore, Barrow agegraphic dark energy exhibits more richer and interesting phenomenology comparing to the standard scenario, and thus it can be a candidate for the
description of nature.\\

\section{Discussion}

One must apply the Barrow entropy for analysing its associated phenomena because of the long-range interaction behaviour of gravity \cite{ref66}. This research is motivated by the Barrow entropy \cite{ref55}, and based on the holographic theory, we introduced a new dark energy model with a time scale as an infrared cut-off. We have considered the IR cut-offs of the model as   the age of the universe and the conformal time. During the cosmic evolution, we examine the nature of equation of state parameter $ \omega_{D}$, deceleration parameter $q$, BADE energy density parameter $\Omega_{D}$ and the squared sound speed $v^{2}_{s}$. We find that both of the presented models are classically unstable except the non-interacting NBADE for $\Delta= 0.3, 0.35$ and 0.4. Moreover, we discuss the importance of the presence of mutual interaction among the dark segments of the universe. We observed that the interacting models proposed here are classically unstable rather than the original TADE model based on the Tsallis entropy \cite{ref63}. To resolve the late-time DE-dominated universe, both the  BADE and NBADE models may be beneficial. Our research points out that for the cosmic evolution, the forecasts of the models are more sensitive to  $\Delta < 2$ rather than $\Delta > 2$. We can achieve this by relating the identical curves, and this influences the primary conditions accepted for graphing the curves. In addition to the squared speed of sound, a performance challenges that the determination of ADE models is the holographic principle and also a complete review on their stability must reflect its non-local characteristics \cite{ref73}. The latter approach of our aim is out with this research and also recognized as an earnest concern for future achievements.

\section{Acknowledgements}
 The author U. K. Sharma thanks
	the IUCAA, Pune, India for awarding the visiting associateship. The authors V. C. Dubey and G. Varshney thank the facility and support provided
	by the GLA University, Mathura, India to lead this research work. The authors are also very much thankful to
the learned referee for his/her constructive suggestions which helped to improve the quality of paper in present form.

\end{document}